# Title: Experimental demonstration of an integrated on-chip p-bit core utilizing stochastic Magnetic Tunnel Junctions and 2D-$MoS_2$ FETs

**Authors:** John Daniel, Zheng Sun, Xuejian Zhang, Yuanqiu Tan, Neil Dilley, Zhihong Chen, and Joerg Appenzeller

**Institutions:** Birck Nanotechnology Center, Purdue University, West Lafayette, IN, United States

## Abstract:

Probabilistic computing is a novel computing scheme that offers a more efficient approach than conventional CMOS-based logic in a variety of applications ranging from optimization to Bayesian inference, and invertible Boolean logic. The probabilistic-bit (or p-bit, the base unit of probabilistic computing) is a naturally fluctuating entity that requires *tunable stochasticity*; by coupling low-barrier stochastic Magnetic Tunnel Junctions (MTJs) with a transistor circuit, a compact implementation is achieved. In this work, through integrating stochastic MTJs with 2D-$MoS_2$ FETs, the first *on-chip* realization of a key p-bit building block displaying voltage-controllable stochasticity is demonstrated. In addition, supported by circuit simulations, this work provides a careful analysis of the three transistor-one magnetic tunnel junction (3T-1MTJ) p-bit design, evaluating how the characteristics of each component influence the overall p-bit output. This understanding of the interplay between the characteristics of the transistors and the MTJ is vital for the construction of a fully functioning p-bit, making the design rules presented in this article key for future experimental implementations of scaled on-chip p-bit networks.

## 1 – Introduction

Computing is at a crossroads: just as the transistor-scaling driven by Moore's Law has afforded improvements in conventional CMOS-based computing performance, there is an inevitable slowing down due to fundamental device limits[1]. Furthermore, the inherently deterministic nature of conventional computing makes the current CMOS model unsuitable for contending with the continued future growth of applications such as in neuromorphic computing and Artificial Intelligence (AI)[2].

A superior approach is that of probabilistic computing. In probabilistic computing, the key component is the probabilistic bit (or p-bit), a unit that fluctuates randomly, but controllably, between 0 and 1[3]. Indeed, a network of such p-bits can leverage their stochastic nature to function as efficient hardware accelerators for solving complex problems that are themselves inherently probabilistic. These problems, which lie at the core of many real-world machine learning applications and algorithms of AI, range in nature from combinatorial optimization problems (such as integer factorization) to recognition and classification[4–16].

At its core, a p-bit requires a tunable stochastic element. While it should be noted that this can be implemented with standard CMOS technology[17–19] and a significant device overhead, the resulting p-bit suffers from a large areal and energy footprint, as well as not offering true randomness[20].

An ultra-compact approach for tunable randomness that yields the desired sigmoidal-shaped input/output characteristics, which is scalable and energy-efficient, is achieved by exploiting the physics of low-barrier fluctuating nanomagnets when coupled with existing Magnetic Tunnel Junction (MTJ) technology. Such p-bit implementations using stochastic MTJs have been shown[21–25], but as yet, the proof-of-concept implementations have required field-programmable gate arrays (FPGAs) or external circuitry, with orders of magnitude more transistors involved than needed in the 3T-1MTJ p-bit design explored in this work.

In this work, the first experimental on-chip demonstration of the core of a p-bit, exhibiting tunable stochasticity, is reported. Using a variation of the 3T-1MTJ design proposed by Camsari et al.[26], a stochastic MTJ is integrated with a high-performance $MoS_2$ transistor next to each other on the same chip, experimentally showing for the first time the desired gate-controlled fluctuations at room temperature. Moreover, this article elucidates the impact and interaction of the various critical device characteristics shown in Figure 1(a), including that of the i) MTJ, ii) the transistor that is part of the p-bit core and iii) the inverter (see Figure 1(a)). It is found that – against common wisdom – a large Tunnel Magnetoresistance (TMR) is not the best choice for p-bits; telegraphic fluctuations are highly undesirable and are a sign of a slow device; matching of the MTJ resistance and the transistor characteristics is crucial; and an ideal inverter with a large gain is incompatible with the desired p-bit operation.

In detail this article is organized as follows: Section 2 briefly introduces the stochastic MTJ, the key element of the p-bit. Next, Section 3 shows the actual hardware demonstration of the p-bit core, focusing on the matching conditions between the MTJ and the transistor. This is followed by a detailed discussion of the impact of Tunneling Magnetoresistance (TMR) and the distribution of states on the p-bit performance in Section 4. Last, in Section 5, the impact of the inverter characteristics is discussed.

## 2 – Implementing probabilistic bits (p-bits) with stochastic MTJs

At its core, a Magnetic Tunnel Junction (MTJ) consists of two ferromagnetic layers separated by an ultrathin insulating layer (Figure 1(b)). The "fixed" layer, which has the stronger magnetic moment, is used as the reference for the "free" layer, whose magnetic moment is more susceptible to being switched. Important MTJ parameters are the Tunnel Magnetoresistance (TMR), that describes the difference in resistance between the parallel (P) and antiparallel (AP) arrangement of the two magnetic layers, and the energy barrier of the free layer, $E_B$, that needs to be overcome to toggle between the two resistance states[27–29].

For stable MTJs, such as those used in spin-transfer torque magnetic random access memory (STT-MRAM) applications[30], energy barriers are large and when the resistance is measured as an external magnetic field is swept, the resulting minor loop exhibits deterministic switching of the free layer. Figure 1(c) shows an example minor loop of a fabricated MTJ that was observed to be stable.

If this energy barrier is made smaller, through material changes or shape scaling[31], the ambient thermal energy may be sufficient for the free layer to switch stochastically between the two resistance states (Figure 1(d)). When biased at the center of this window, the signal is shown to be a naturally fluctuating output whereby the time spent in each resistance state (known as the dwell time, $\tau$) may be described by the equation:

$$\tau = \tau_0 e^{E_B/K_B T}, \qquad (1)$$

where $k_B$ is the Boltzmann constant, $T$ is the temperature and $\tau_0$ is the "attempt time", a material-dependent constant that is ~1ns[32]. For in-plane stochastic MTJs, dwell times down to ~5ns have been demonstrated[33,34].

For p-bit applications, this source of natural stochasticity is ideal; by coupling a stochastic MTJ with an access transistor, and including an inverter for amplification, a compact voltage-controlled p-bit design is achieved (Figure 1(a))[26].

The theoretical output from such a p-bit implementation, generated using modified experimental data from stochastic MTJs (Figure 1(e)) and circuit simulations of transistor behavior (Figure 1(f)), is shown *before* (Figure 1(g)) and *after* (Figure 1(h)) the inverter's amplification. (For more details regarding the use of experimental data in the circuit simulations, please see Supplementary Information 1). The core of the p-bit, which includes the stochastic MTJ and NMOS transistor, provides the tunable stochasticity while the inverter provides the thresholding and amplification of the stochastic signal. The resulting sigmoidal output allows for pinning at low- and high-input voltages, while exhibiting the desired output fluctuations in the transition region.

The tunability in the output is controlled through varying the transistor gate voltage ($V_{IN}$), where changes in the relative resistance of the transistor to the MTJ change the voltage at the inverter's input. This voltage is then amplified through the inverter's operation, allowing the output to be pinned to output-low for low $V_{IN}$, and to output-high for high $V_{IN}$. In the middle region, the stochastic resistance fluctuations from the MTJ manifest as tunable random voltage fluctuations in the p-bit output.

This design is discussed further in the following section, which shows the experimental realization of the p-bit core using a stochastic MTJ and a 2D-$MoS_2$ transistor.

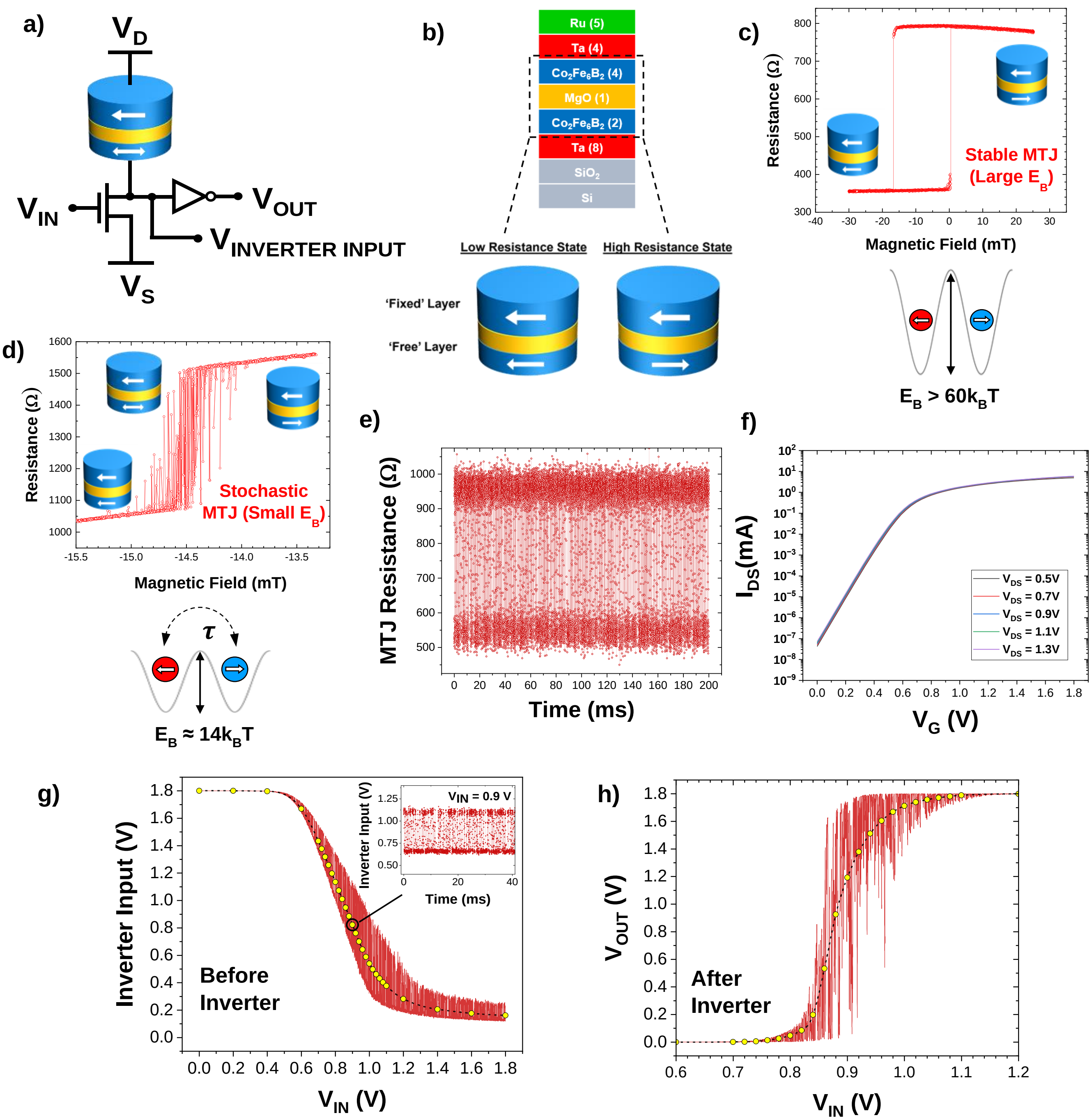

**Figure 1: Implementing p-bits with stochastic MTJs.** a) Schematic of the proposed p-bit design, comprised of a stochastic MTJ connected to the drain of an NMOS transistor, forming the stochastic core of the p-bit, and an inverter, for thresholding and signal amplification. b) Cross-section schematic of a typical MTJ stack, with layer thicknesses denoted in nm, and an explanation of the TMR effect. c) Minor loop of a stable MTJ, showing how the resistance changes deterministically as a function of magnetic field for a large energy-barrier free layer. d) Minor loop of a stochastic MTJ, showing how the resistance fluctuates between the parallel- and anti-parallel state for a small energy-barrier free layer. e) Example time-series resistance data for the fluctuating MTJ and f) transfer characteristics of the transistor used to obtain the example p-bit's output. g) Graph showing the typical p-bit signal before the inverter's operation, as a function of the transistor gate voltage (defined as $V_{IN}$). The average at each point is shown by the dotted line. h) Graph showing the typical output of the full p-bit, $V_{OUT}$, as a function of the input voltage, $V_{IN}$. The time-averaged signal at each input voltage is represented by the dotted line.

## 3 – Experimental demonstration of a stochastic on-chip p-bit core with integrated stochastic MTJs & 2D FETs

For this demonstration, MTJ devices were first fabricated before those devices possessing sufficient TMR for a large read-signal are interconnected with appropriate resistance-matched field effect transistor (FET) devices in a 1T-1MTJ configuration. It is desirable to have the transistor chosen such that the on-state FET resistance is at least two orders of magnitude smaller than the MTJ's low-resistance state, $R_P$, and that the off-state FET resistance is two orders of magnitude larger than the MTJ's high-resistance state, $R_{AP}$, to attain the maximum swing in the output voltage.

Figure 2(a) shows a schematic of the 1T-1MTJ configuration for the on-chip p-bit core. The detailed stack structure for the MTJs used in this demonstration is shown in Figure 2(b). The magnetic layer (CoFeB) thicknesses, were chosen to best yield MTJs with in-plane anisotropy due to two reasons: MTJs with in-plane anisotropies have been shown to be more resistant to Spin Transfer Torque (STT)-pinning[35], and have also shown to fluctuate with time scales that are orders of magnitude faster than perpendicular-anisotropy MTJs [32,34,36].

Figure 2(c) shows an SEM image of an exemplary elliptical nanopillar with the same dimensions as the MTJs used in this demonstration, while Figure 2(d) shows an optical microscope image of a finished MTJ device, along with a tilted-angle false-color SEM image of the MTJ region.

The interdigitated (IDT) monolayer (ML) $MoS_2$ FETs are then fabricated alongside the completed MTJ devices. The cross-section of the FET is shown in Figure 2(e) while an SEM image of a fabricated IDT ML $MoS_2$ FET is shown in Figure 2(f), where a single IDT FET includes 20 sets of source/drain contacts, with $L_{ch}$~150nm and $W_{ch}$~6.5µm, for a total effective channel width of 130µm. ML $MoS_2$ is chosen as the channel material of the drive transistor due to the low thermal budget fabrication process (to help preserve the performance of the fabricated stochastic MTJs, which suffer shorting in the $SiO_2$ isolation layer for temperatures above ~400˚C), low contact resistance[37], the large bandgap (1.8eV), high on-state performance of scaled 2D-$MoS_2$ FETs[38] and good electrostatic control achievable with ML $MoS_2$. Note that the ultimate p-bit implementation would involve combining advanced CMOS circuitry with unstable MTJs (rather than using MTJs in an MRAM array structure as nonvolatile memory elements).

Figure 3(a) shows the minor loop of the stochastic MTJ used in the integrated p-bit. The dashed line at -16mT indicates the 50-50 point at which the device "spends" an equal amount of time in the AP- and P-state. All further measurements for this device are performed at this 50-50 point to ensure the MTJ's resistance output (Figure 3(b)) is truly random. As this is an intrinsically Poisson process, fitting the histograms of the AP- and P-state dwell times (Figure 3(c)) with an exponential envelope yields the average dwell time in each state ($\tau_{AP}$ and $\tau_P$, respectively)[20,39]. The dwell time of this device, a quantity that determines the speed at which a p-bit may operate, is calculated as the harmonic mean of $\tau_{AP}$ and $\tau_P$ and is 695ms (Details on the dwell time extraction and the quality of randomness can be found in Supplementary Information 2).

The transfer characteristics of 24 as-fabricated IDT ML $MoS_2$ FETs are seen in Figure 3(d), showing a narrow variation in the threshold voltage, while the benefits of the IDT structure are seen in the high current levels and on/off ratios. The on-current level is around 0.6mA at $V_{DS}$ = 0.1V and the on/off ratio is around ~ $10^{10}$, with a minimum subthreshold slope (SS) around 94 mV/dec. Note that the scaled devices operate at gate voltages on the order of ~1V, which is critical for the ultimate p-bit implementation to ensure that $V_{IN}$ and $V_{OUT}$ are identical.

Following the characterization of devices, a Ti/Au interconnect is fabricated between the MTJ- and $MoS_2$-FET pair observed to have the best resistance match and stochastic signal. It is observed, however, that after the integration of MTJ and FETs, there is a degradation in the transistor performance, as shown in Figure 3(e), including degraded on-off ratio and SS. This is not a result of connecting the FET with the MTJ, but likely due to process-induced trap charges in the $HfO_2$ gate oxide that produced an aging effect, whereby the FET characteristics were observed to degrade over time for this device[40].

A circuit schematic of this 1T-1MTJ p-bit core is shown in Figure 3(f), while an optical microscope image of the finished device is shown in Figure 3(g). Figure 3(h) shows the output, $V_{INVERTER\ INPUT}$, as a function of the input (FET gate) voltage, $V_{IN}$. $V_D$ = 200mV was used to avoid excessive stress to the MgO barrier and to prevent the damage to the MTJ observed at larger current densities. (To better understand the choice of $V_D$ and the impact of large current densities through the MTJ, see Supplementary Information 3).

For this measurement, the MTJ is biased at its 50-50 point (as seen in Figure 3(a)) and $V_{INVERTER\ INPUT}$ is measured 200 times at each input voltage value, $V_{IN}$, to demonstrate the impact of the stochastic fluctuations on the p-bit core's output.

To compensate for the transistor degradation in the integrated p-bit core, $V_{IN}$ had to be significantly increased, which will not be required in a further optimized p-bit implementation. At large negative $V_{IN}$, when the transistor is in its highly resistive OFF-state, the potential at $V_{INVERTER\ INPUT}$ is close to $V_D$. Increasing $V_{IN}$ yields a decrease in the transistor's resistance, resulting in a reduction in $V_{INVERTER\ INPUT}$ as the transistor approaches its threshold voltage, $V_{TH}$.

For this device, the leftward shift of the degraded transistor's threshold voltage, $V_{TH}$, results in a leftward shift of the overall sigmoid while the degradation in the transistor's off-state resistance (shown in Figure 3(e)) results in the output not being fully pinned to $V_D$ (see Supplementary Information 5 for off-chip p-bit core implementations with better resistance-matching and better $V_{IN}$-$V_{OUT}$ matching between the constituent MTJ-FET pair, illustrating that the non-idealities in the on-chip demonstration discussed here are a result of process modules and not a fundamental issue).

The impact of the MTJ's fluctuations also becomes increasingly clear in the p-bit core output as $V_{IN}$ is increased, with the magnitude of fluctuations observed at a maximum when the resistances of the transistor and the MTJ are approximately equal, and an equal voltage is dropped across both components. The red inset in Figure 3(h) reveals a significant voltage drift in the output due to charge traps from the degradation of the transistor gate oxide and its impact on the subthreshold slope.

A further increase in $V_{IN}$ to the transistor's ON-state, where the resistance of the transistor is less than that of the MTJ, sees the output approach 0V. The output here still shows the fluctuations from the MTJ, but at a much smaller scale (green inset, Figure 3(h)). This is beneficial as any STT-pinning effects from the large currents at this input voltage, that could act to potentially bias the 50-50 fluctuations of the MTJ, do not significantly impact the output of the p-bit core (Supplementary Information 3 shows how large current densities through the MTJ can result in STT-pinning).

In this way, this demonstration of a scaled on-chip p-bit core is shown to produce the desired sigmoidal output with the *tunable* stochasticity that is required for probabilistic computing. A desirable feature of the sigmoid is that it is centered around $V_{IN} = V_D/2$, such that $V_{IN}$ and $V_{OUT}$ may be of similar scales, and the output of one p-bit may be fed into the input of another p-bit to create correlated p-bit networks. This may be achieved by implementing a dual-gated transistor design, whereby the threshold voltage may be shifted to the desired region through the application of an additional top-gate voltage (demonstrated in Supplementary Information 4).

This demonstration also illustrates the impact the transistor has on the p-bit's output. For example, the subthreshold slope (SS) determines the steepness of the sigmoid (a steeper SS would yield a steeper sigmoid), and how well the transistor is resistance-matched with the MTJ impacts the $V_D$ range over which the output sigmoid spans and if the output can be pinned. Moreover, the location of the threshold voltage is critical in determining the centroid of the overall sigmoid (as shown in Supplementary Information 5).

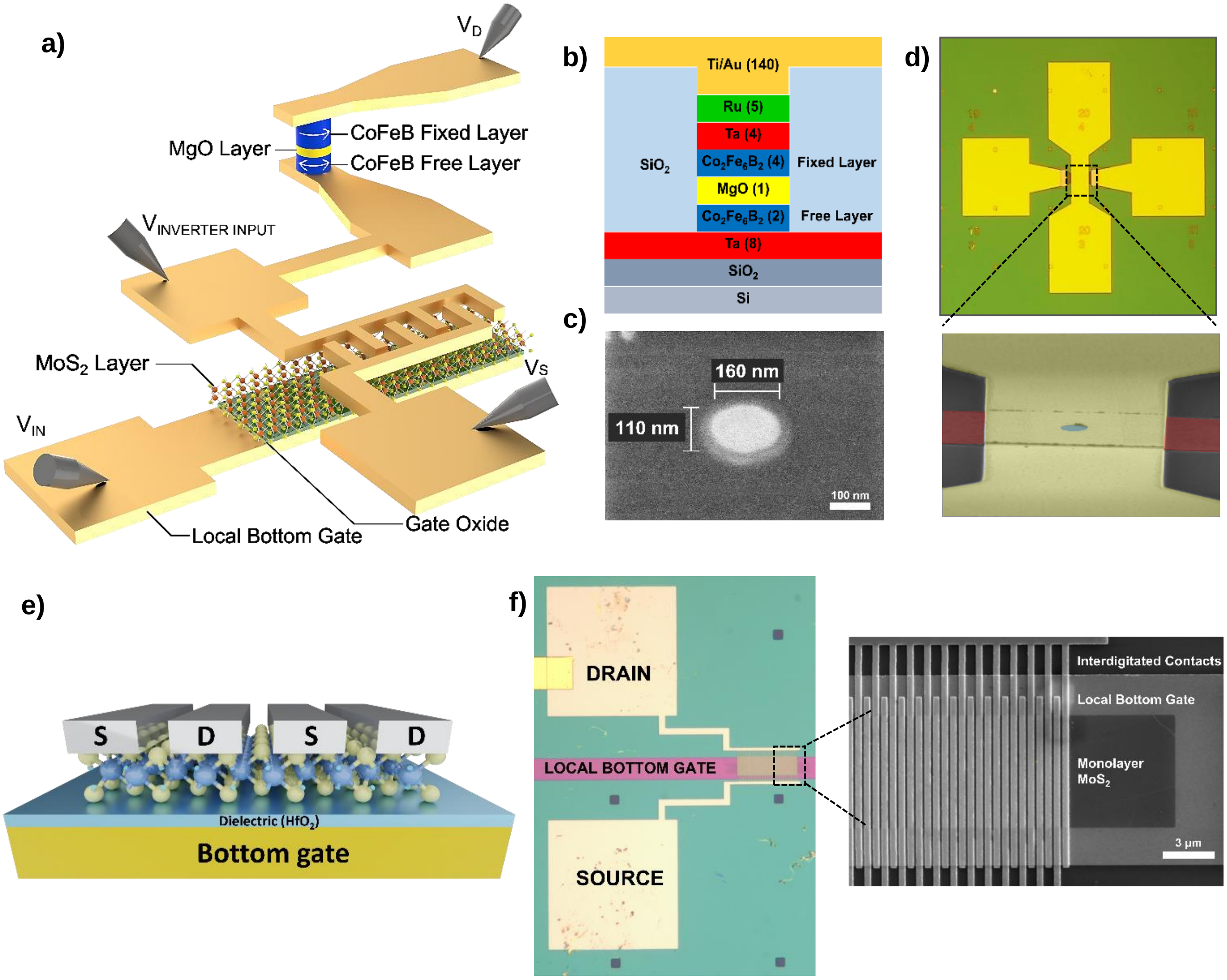

**Figure 2: Fabricating an on-chip p-bit.** a) 3D schematic of the proposed design for an on-chip p-bit core, using a stochastic MTJ and 2D $MoS_2$ FET. b) Side cross-section view of the MTJ stack, with the fixed- and free-layer denoted. c) Top SEM view of an example MTJ pillar of the same nominal dimensions as the stochastic MTJ in the integrated on-chip device. d) Optical microscope and (false-color) tilted-SEM images of an example finished MTJ device. e) Cross-section schematic of the 2D $MoS_2$ FET. f) Optical microscope and SEM-images of the 2D FET, showing the interdigitated contacts that are used to attain high-current drives.

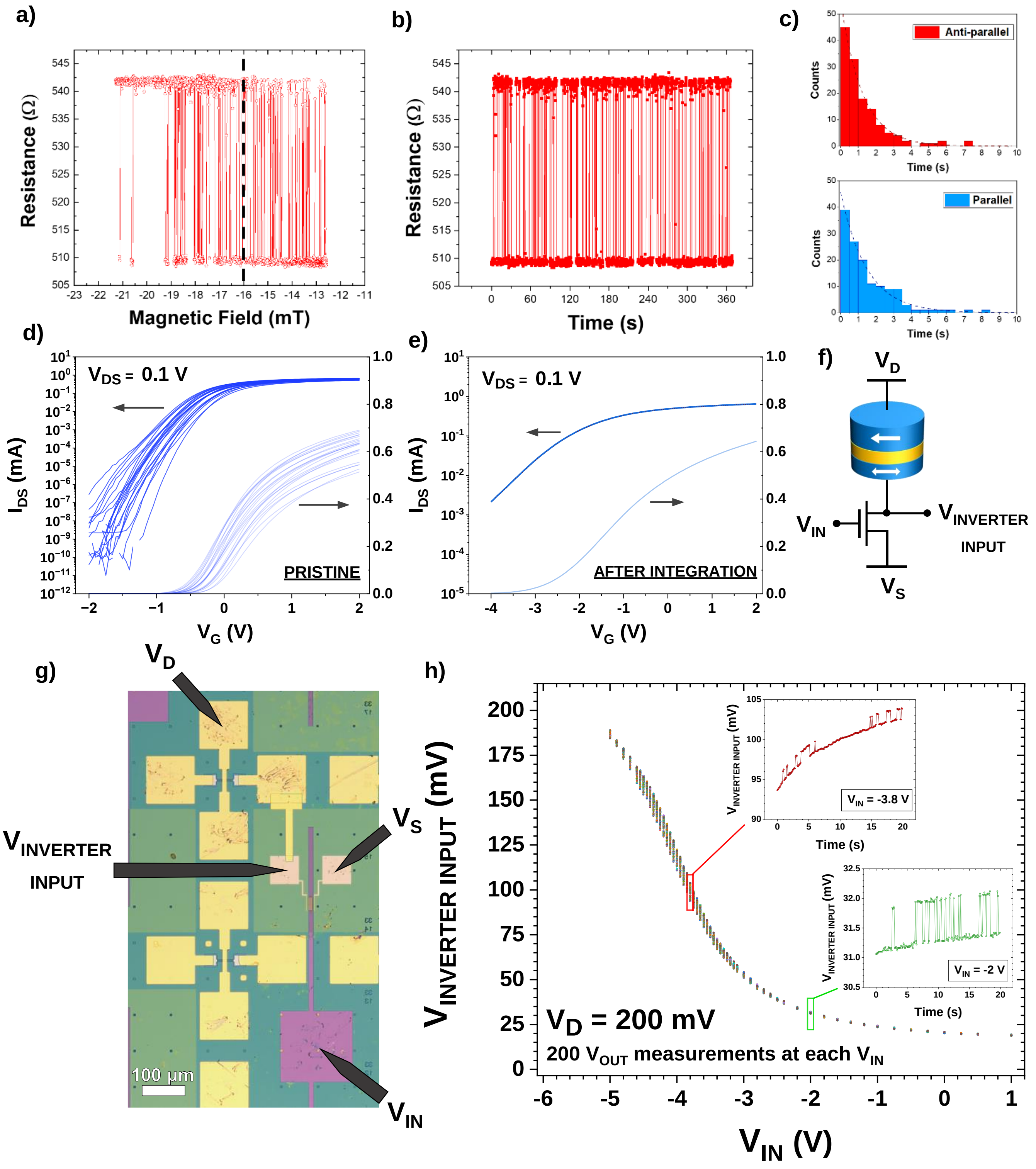

**Figure 3: Characterization & measurement of the integrated on-chip p-bit core.** a) Minor loop of the stochastic MTJ used in the p-bit, with the 50-50 point indicated by the dashed line. b) Graph showing the resistance fluctuations of the stochastic MTJ when biased at the 50-50 point. c) Histogram data for the anti-parallel- and parallel-state dwell times, used to extract the mean dwell time in each state. d) Graph showing the device characteristics of the pristine $MoS_2$ FETs using the interdigitated contacts design, before being connected to the stochastic MTJs. e) Graph showing the device performance of the FET used in the on-chip p-bit core. The degradation was observed after integrating the FET with the stochastic MTJ device. f) Schematic of the design for the on-chip p-bit core demonstration. g) Optical microscope image showing the integrated on-chip device and the probes used for measurement. h) Graph showing the operation of a truly on-chip p-bit core, with an output ($V_{INVERTER\ INPUT}$) that exhibits stochastic fluctuations that is tunable with the modulation of the FET gate voltage ($V_{IN}$).

## 4 – Influence of MTJ characteristics on the p-bit output

To study the impact of an MTJ's characteristics on a p-bit's output, experimental data from stochastic MTJs are used as input for circuit simulations, conducted using the Spectre Simulation Platform. A 3T-1MTJ model of the p-bit is used (Figure 4(a)), with additional bias points available at the body bias for the NMOS and PMOS transistors of the inverter for tuning of inverter characteristics (Further information about data handling, and the transistors that are part of the p-bit circuitry, is provided in Supplementary Information 1).

Two key properties of an MTJ are investigated: the MTJ's TMR and the MTJ's distribution of resistance states. An ideal p-bit output in a 3T-1MTJ configuration would be a smooth sigmoidal function with a wide region of fluctuations, at the center of which are rail-to-rail fluctuations that could be used to drive other p-bits in a network of such devices.

Figure 4(b) shows the p-bit output for three MTJs fluctuating at the same frequency, but with TMR ratios scaled to different values (Supplementary Information 6 describes how this TMR-scaling was performed using actual measured MTJ fluctuations). The dotted line shows the time-averaged $V_{OUT}$ at each $V_{IN}$, while the shaded background shows the instantaneous output as $V_{IN}$ is swept linearly from 0 to 1.8V.

The largest TMR device (300%, blue) has the widest stochastic region and rail to rail fluctuations, but also shows a plateau in the time-averaged curve. These plateaus, or the pinning of the output over a range of input voltages, are non-ideal for concatenation purposes as it reduces the tunability of an individual p-bit's fluctuations with changes in its input from other p-bits in the network.

In contrast, the smallest TMR device (15%, black) has no plateaus but has a narrow range over which the fluctuations are visible. This is undesirable as it limits the $V_{IN}$ range in which usable fluctuations are observed, with the p-bit output primarily in the output-low or output-high state. To understand this behavior, consider Figure 4(c).

Figure 4(c) shows, for increasing $V_{IN}$ applied to the transistor's gate, the distributions of values at the inverter's input for each of the p-bits made with MTJs of differing TMRs, along with the voltage transfer curve (VTC) of the inverter (overlaid in green). The largest TMR device (300%, blue), with the largest resistance fluctuation, has the widest spread of values for Inverter Input, while the smallest TMR device (15%, black) has the narrowest distribution (Supplementary Information 7 provides further explanation of these voltage distributions).

For $V_{IN}$ = 0.8V (left graph), the value at the inverter's input is centered around $V_{INVERTER\ INPUT} \approx$ 1.2V, such that the $V_{OUT}$ is within output-low, i.e. close to zero, on the VTC for both the 15% and 80% TMR. However, the 300%-TMR device has a sufficient number of states in the bottom arm of its $V_{INVERTER\ INPUT}$ distribution (blue) that is in-between the noise margin regions of the inverter's VTC, such that the average $V_{OUT}$ for the 300%-TMR device is shifted to a larger value of ~320mV (Figure 4(b)).

As $V_{IN}$ is increased, the transistor connected to the stochastic MTJ becomes more conducting, and the center of the distributions shift to smaller $V_{INVERTER\ INPUT}$ values. For $V_{IN}$ = 0.98V (right graph in Figure 4(c)), the 15%-TMR device (black) has inverter input values such that it interacts primarily with the output-high section of the VTC, giving an average $V_{OUT}$ that is pinned close to 1.7V (Figure 4(b)).

In contrast, the 300%-TMR device has a larger range of inverter input values that spans between the noise margin regions of the inverter. This results in the plateau effect, where changing the input voltage does not yield a meaningful change in average $V_{OUT}$ as the TMR is large enough for the distribution to span both the output-high and output-low regions of the VTC for a range of $V_{IN,}$ and inverter input, values.

To summarize, the smaller the TMR, the smaller the section of the VTC that is 'sampled' by the inverter input distribution, and the smaller the range of $V_{OUT}$ over which the values are averaged. This results in a smoother averaged output that is more sigmoidal and less prone to plateauing. However, the $V_{IN}$ range over which the stochastic fluctuations are observed is small, limited to the range between the output-high and output-low regions of the VTC, where the gain is non-zero. This means that for a small TMR device, rail-to-rail fluctuations are not observed at all. Although it has been shown that rail-to-rail fluctuations are not necessary for the entire fluctuating range[26], the diminished output fluctuation range would make it difficult to form networks with small-TMR p-bits due to the insufficient voltage drive it would provide to the next p-bit. A large TMR-device is good for attaining rail-to-rail output voltages, such as at $V_{IN}$ = 0.9V (middle graph) where the 300%-TMR device shows an output spanning 0 to 1.8V, but is prone to the plateauing effect if the device's inverter input distribution spans the output-high and output-low regions of the VTC for an extended range of $V_{IN}$ values.

This is a key finding: for a given inverter, the TMR should not be too high such that it spans the output-low and output-high regions of the inverter for a large $V_{IN}$ range. A 'perfect' inverter that has an infinite gain would be undesirable for p-bit applications, as even an MTJ with a small TMR would have a step-like plateau in the output.

Another characteristic that affects the p-bit's output is the MTJ's distribution of states. An MTJ with a very bimodal distribution is more prone to plateaus in the output, especially if the TMR is large enough for the fluctuations in the inverter's input to sample both output-high and output-low regions of the VTC. In contrast, an MTJ with a very continuous distribution, with the ideal being a uniform distribution between $R_P$ and $R_{AP}$, would sample each value of the VTC equally and would give a much smoother sigmoidal output[41].

A further key finding of this work is that there appears to be a correlation between the distribution of resistance states and the speed at which these in-plane MTJs fluctuate. To quantify how bimodal a MTJ's resistance fluctuations are, a new figure-of-merit, the 'distribution factor', is introduced. Using the normalized resistance output of an MTJ, histograms are created where the counts in the 8 'edge' state bins are divided by the 8 'middle' state bins. For statistical significance, the total number of data points is the same in each data set. Figures 5(a) and 5(b) show this process for two MTJs of different dwell times ($\tau$ = 117µs and $\tau$ = 27ms, respectively).

Figure 5(c) shows this distribution factor calculated for 23 stochastic MTJs, made with the same stack material, with dwell times spanning orders of magnitude (Supplementary Information 8 explains in greater detail why it is meaningful and justified to use the distribution factor as a key metric).

It is observed that the faster the MTJ fluctuates, the more middle states there are in the resistance distribution, and the less bimodal the distribution is. The dotted line is a guide to the eye which suggests that *for this material stack*, a uniform distribution with an equal edge- and middle-state counts would be achieved for MTJs with fluctuations in the tens of ns regime. This correlation suggests that a faster device, with a more continuous distribution, would yield a smoother sigmoidal output.

This is tested with the two devices of different dwell times, $\tau$ = 117μs and $\tau$ = 27ms, that are scaled to the same TMR, and using the same inverter. Figure 5(d) shows that the faster device (117μs, red), which has a smaller distribution factor and is less bimodal, yields a p-bit output that is more ideal than the output from the slower device (27ms, orange) which suffers from the plateauing effect described previously.

This is another key finding in that a faster MTJ has a two-fold advantage: firstly, the faster the fluctuation and speed of random number generation, the faster the p-bit may operate asynchronously, and secondly, the faster the MTJ, the more uniform the distribution of states is observed to be, and the more ideal the p-bit's output is. Thus, it is this interplay of the MTJ's TMR and the distribution of states, along with the inverter's properties, that can determine how ideal a p-bit's output is.

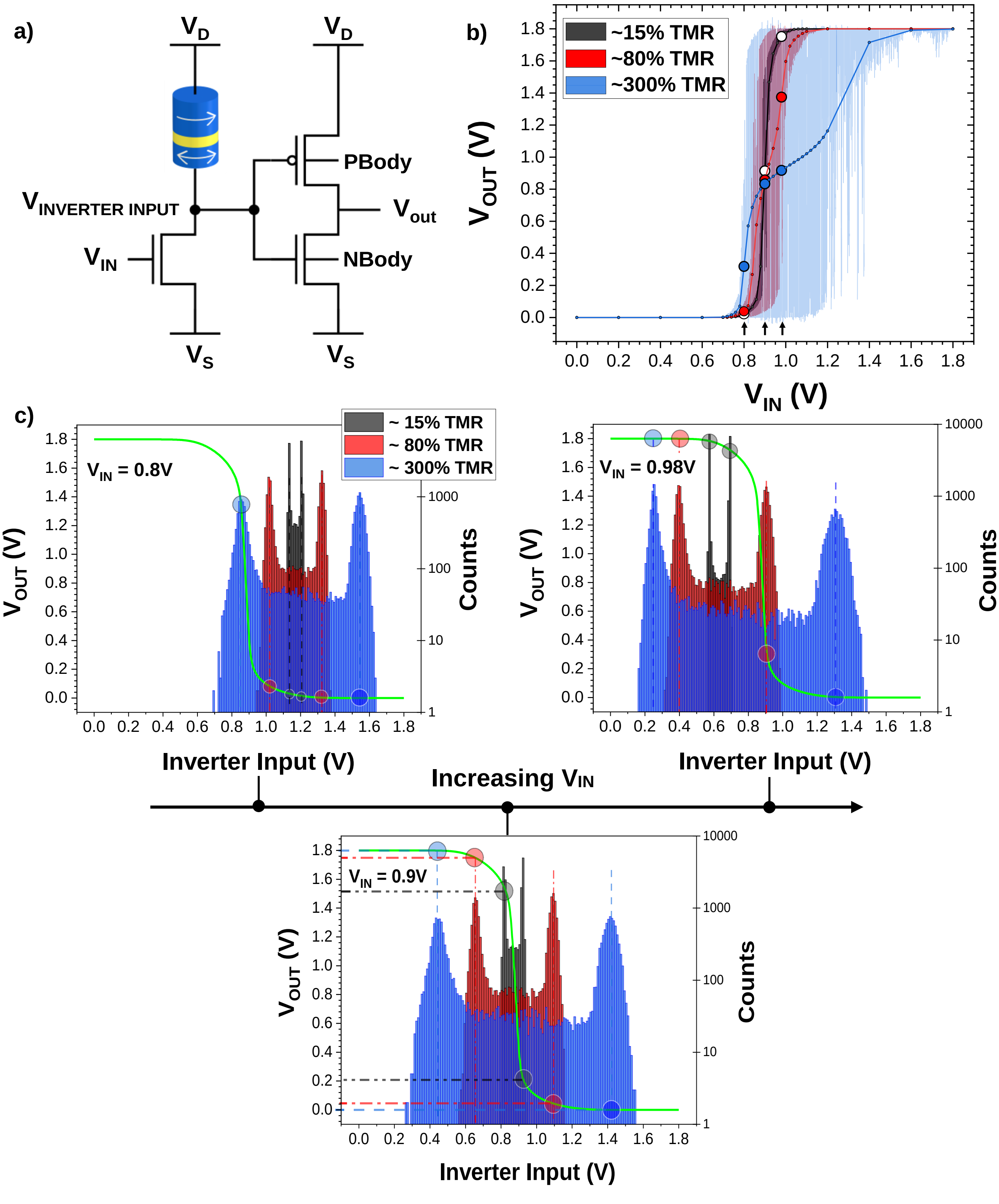

**Figure 4: Influence of the MTJ's TMR on the p-bit output.** a) Diagram of the p-bit circuit implemented in Cadence software for circuit simulations. b) Graph showing the outputs of p-bits made with MTJs of differing TMRs; an MTJ with too large a TMR is likely to cause undesirable plateaus in the p-bit output. c) Graph showing the distribution of voltages at the inverter's input (histogram data, right axis) for the different TMR devices, and how this interacts with the voltage transfer curve of the inverter (green curve, left axis), to produce the averaged $V_{OUT}$ signal for $V_{IN}$ values of 0.8V (left graph), 0.9V (middle graph) and 0.98V (right graph).

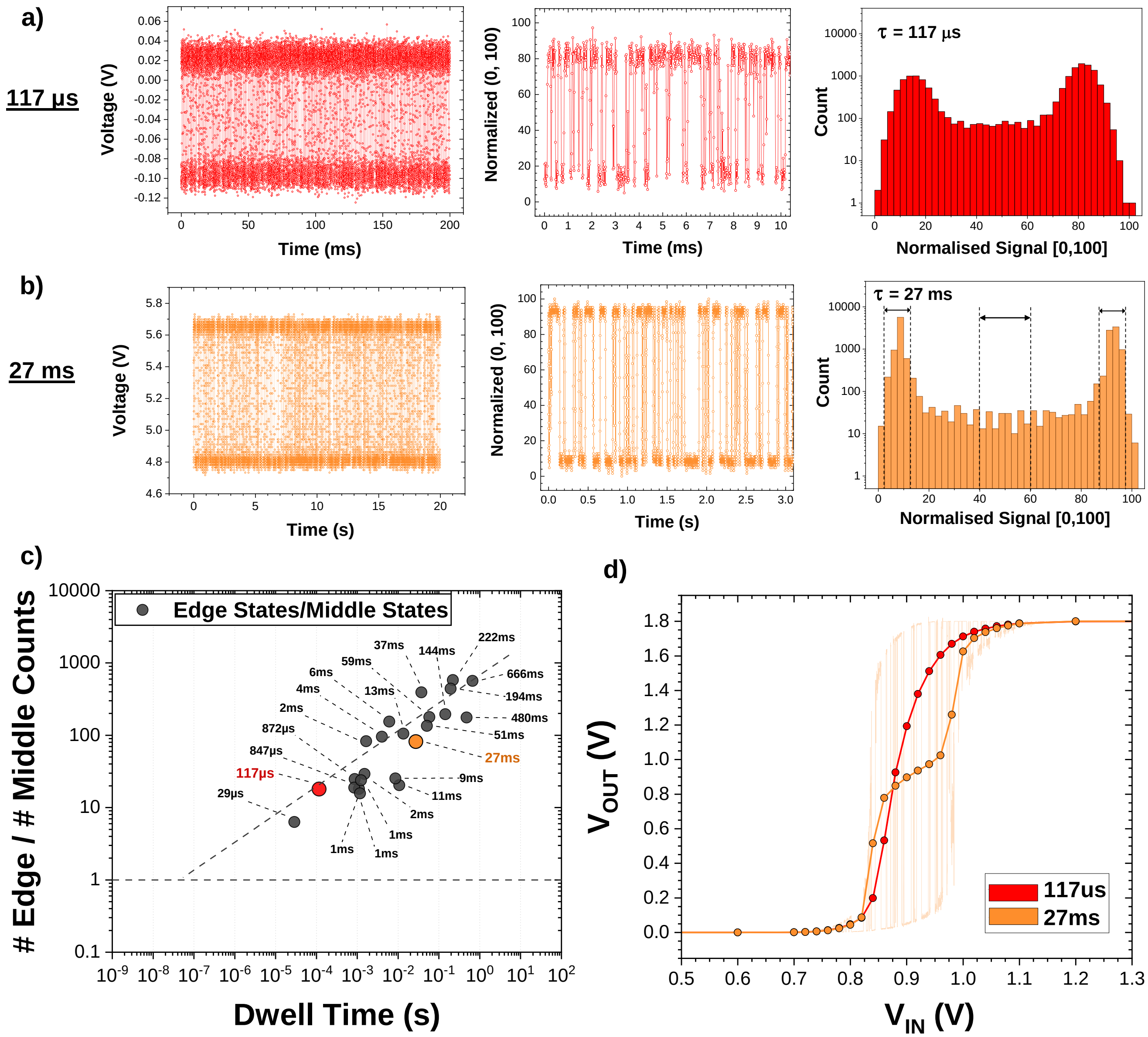

**Figure 5: Resistance distributions of stochastic MTJ devices.** a) Data for a stochastic device with a dwell time of 117μs, showing the raw signal as measured at the oscilloscope (left), the normalized resistance data, zoomed in to show the fluctuations are well-sampled (middle), that are used to create the device's normalized resistance histogram (right). b) Stochastic device data, as described previously, for a device of a slower dwell time of 27ms, showing correspondingly fewer 'middle states' in the normalized resistance histogram. c) Graph showing the "distribution factor" of 23 stochastic MTJ devices, a quantity defined as the number of 'edge states' divided by the number of 'middle states' in an MTJ's normalized resistance histogram, as a function of their measured dwell times. The dotted line is a guide to the eye, showing the predicted behavior for even faster devices made with this material stack. d) Graph showing the averaged output for two p-bits that use the same inverter and transistor, but with MTJs of different dwell times and distributions.

## 5 – Influence of inverter characteristics on the p-bit output

The inverter also offers a degree of control over the p-bit's output. Figure 6(a) shows the voltage transfer curve (VTC) for two inverters: one without applied body bias, called "pristine" (black curve), and the other which has been tuned, through application of a positive body bias to the NMOS-FET, to have a smaller gain (red curve).

Using the same MTJ and transistor, Figure 6(b) shows the impact of this inverter tuning on a p-bit's output: the tuned inverter (red), with the smaller gain, shows a smoother sigmoid while the pristine inverter, with the larger gain, shows a more pronounced undesirable plateau in the output. This is because for a given MTJ with a bimodal distribution, the distribution of voltages at the inverter's input is less likely to span the output-low and output-high regions for an extended range of $V_{IN}$ (the cause of the undesirable plateaus) if the VTC is shallower and the gain is small.

However, the tuned inverter also suffers from a degradation in the noise margin, seen in Figure 6(a), which decreases the size of the p-bit's output fluctuation range. This is because the body bias at the NMOS transistor shifts its threshold voltage, lowering the channel resistance and making it harder to pin to output-high, $V_D$, for large $V_{IN}$.

This issue could be mitigated by using a more aggressively scaled technology node for the inverter than the 180nm-node used here. A 14nm-ultrascaled Fin-FET inverter (as used in previous p-bit simulation work[26,41,42]), which provides a more piecewise-linear VTC that offers a lower gain (for a smoother sigmoidal output), and a wide-noise margin to pin the output to $V_D$ at high input voltages, would be desirable.

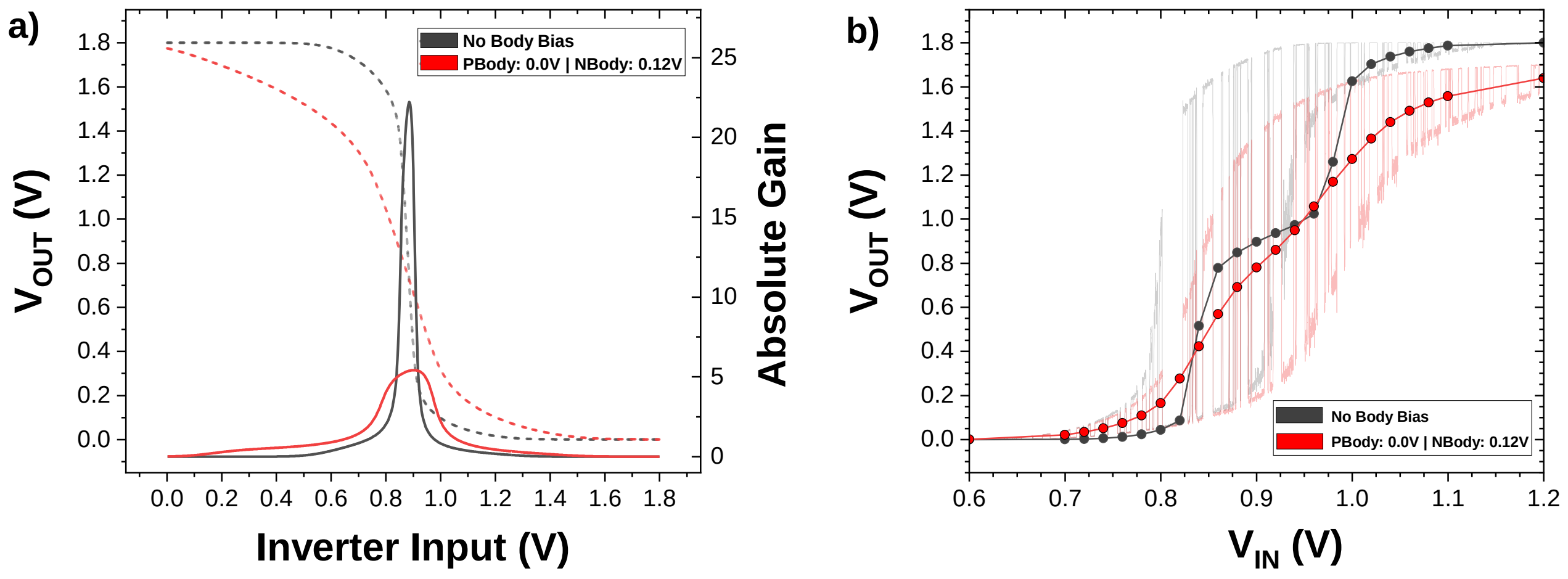

**Figure 6: Influence of the inverter's characteristics on the p-bit output.** a) Graph showing the voltage transfer curve (dotted line, left axis) and the absolute gain curve (solid line, right axis) for an untuned inverter (black) and an inverter that is tuned (red) to artificially lower the gain (by application of a body bias to the NMOS transistor in the inverter). b) The p-bit output, $V_{OUT}$, as a function of the input voltage, $V_{IN}$, for the untuned (black) and lower-gain inverter (red).

**6 – Conclusion**

In this work, the first experimental realization of an *on-chip* p-bit core is demonstrated, using a stochastic in-plane MTJ integrated with a 2D-$MoS_2$ transistor in a 1T-1MTJ structure. Through experimental demonstration and circuit simulations, it is shown how each component of the p-bit influences the overall output.

For the transistor, a good resistance-match with the MTJ and a threshold voltage close to $V_D/2$ is required to achieve a well-centered sigmoid that spans the full range of $V_D$ and is suitable for inverter amplification.

For the stochastic MTJ, too large a TMR can cause plateaus in the inverter's average output, while too small TMR gives an insufficient $V_{IN}$ range over which the usable fluctuations in $V_{OUT}$ are observed. Additionally, it is found that the speed at which the MTJ fluctuates is crucial to the p-bit's output: a faster MTJ is observed to have a more uniform distribution (with more middle states between $R_P$ and $R_{AP}$ edge states), and for a given inverter, this results in a smoother $V_{OUT}$ sigmoid with less plateauing. A faster MTJ is also beneficial when concatenating p-bits to form a p-bit network, whereby the speed of the MTJs used can determine the speed of asynchronous operation.

For the inverter, the large gain and the steep VTC associated with the conventional 180nm-node technology used in the simulations was found to be more likely to yield undesirable plateaus in the p-bit output. A smaller gain inverter, with a piecewise linear VTC that maintains a wide noise margin in the input-low and input-high regions, achievable with a more scaled process, is desirable for p-bit applications.

These observations highlight how each component is crucial in determining the quality of the p-bit's output and seek to provide design insights that can contribute towards the future goal of fully scaled on-chip p-bit networks.

## Methods:

### MTJ Fabrication

MTJ films are deposited using DC/RF sputtering on thermally oxidized Si substrates and, from the bottom, are Ta(8 nm)/CoFeB(2 nm)/MgO(1 nm)/CoFeB(4 nm)/Ta(4 nm)/Ru(5 nm).

These stacks are patterned into elliptical nanopillars using e-beam lithography and Ar-ion beam etching. Amorphous SiO2 is then deposited, to electrically insulate the bottom contact channel, with the etch hard mask in place as part of a self-aligned process. The hard masks are then removed using NMP-based solvent, after which the MTJs are annealed at 300˚C for 10 minutes to improve the TMR of the finished devices[43]. After the annealing procedure, the top contacts are defined using e-beam lithography, with e-beam evaporation used to deposit Ti/Au (20/140 nm) electrodes to enable electrical measurements across the MTJ.

### 2D FET Fabrication

The bottom gate electrode structure is made of a Cr (2nm)/Au(13nm) metal stack followed by 5.5nm $HfO_2$ gate oxide. The $HfO_2$ is deposited by an atomic layer deposition (ALD) system at 90ºC. Then the ML $MoS_2$ flakes are wet transferred from the original $Si/SiO_2$ growth substrate onto the bottom gate electrodes and then vacuum annealed at a pressure of ~$5\times10^{-8}$ torr at 200ºC for 2 hours. After vacuum annealing, the flakes are etched into a stripe before the interdigitated source/drain contacts are defined by electron beam lithography (EBL) and Ni (70nm) is deposited as the contact metal by electron beam evaporation.

## Data Availability

The data that supports the figures and the findings within this paper are available from the corresponding author upon reasonable request.

## Acknowledgements:

The authors thank Prof. K. Camsari for the many helpful discussions and for their invaluable insight. This work was supported by the National Science Foundation (NSF) through Award Number 2106501.

## Author Contributions

J. A. and Z. C. conceived of, and supervised, the project. N.D. provided film-level analysis of the Magnetic Tunnel Junction (MTJ) stacks from which J.D. fabricated the stochastic MTJ devices. J.D. and Y.T characterized the stochastic MTJ devices. Z.S. fabricated and characterized the 2D-MoS2 FET devices. J.D. and Z.S. fabricated and measured the integrated on-chip device. X.Z. performed the circuit simulations and X.Z., J.D., J.A. and Z.C. analyzed the results.

J.D. and J.A. wrote the manuscript, with contributions from Z.S and X.Z. All the authors discussed the data and resulting outcomes.

## Competing interests

The authors declare no competing interests.

**Experimental demonstration of an integrated on-chip p-bit core utilizing stochastic Magnetic Tunnel Junctions and 2D-$MoS_2$ FETs – Supplementary Information**

## Supplementary Information 1 – Simulation Details

The simulation work presented in this paper is carried out on the Virtuoso Cadence EDA tool, using the Spectre Circuit Simulator. Both design and simulation of the p-bit are implemented using an open-source, and open-access, 180nm design kit from North Carolina State University (NCSU).

The design of p-bit consists of an MTJ connected to the drain of an NMOS transistor, forming the core tunable stochastic unit, and an inverter, which provides thresholding and amplification of the signal from the stochastic unit.

For simulation purposes, the MTJ is implemented by using a variable resistor, through which experimental data from real stochastic MTJs is fed. To prepare the data for input, the raw measurement from the oscilloscope is taken and normalized between the range [0,100], before being scaled to a certain resistance range. For the simulation work involving the comparison of different TMR coefficients, data from the same device is used and scaled to different TMR ranges. For the comparison between devices of different distributions and dwell times, data from different MTJs were scaled to the same ~80% TMR. Each resistance data set consists of 16,000 data points, with the simulator's timestep chosen such that it matches the timestep of the sampled MTJ data. This ensures the preservation of the distribution of the real MTJ's resistances. This is because data is fed into the simulator using the piece-wise-linear (PWL) algorithm that would linearly interpolate between two adjacent datapoints if the simulator's timestep is smaller than that of the sampled MTJ's resistance data. As shown in Supplementary Figure 1(a), such interpolation can alter the statistical distribution of the MTJ data and potentially impact the validity of comparison between the different MTJ p-bits.

In series with the MTJ is a NMOS transistor that is 11.25μm wide and 180nm long, designed to be resistance-matched with the MTJ in the simulation such that the full span of $V_D$ at the inverter's input is attained.

The inverter is comprised of a PMOS transistor that is 810nm wide and an NMOS transistor that is 270nm wide, with both being 180nm long. For simulations that do not bias the inverter, the body bias of the inverter's PMOS transistor is set to 1.8V ($V_{DD}$) and the inverter's NMOS transistor to 0.0V ($V_{SS}$).

The example outputs in Supplementary Figure 1(b) and 1(c) (reproduced from Figure 1 of the main text) are obtained by sweeping $V_{IN}$ and sampling the voltage levels *before* the inverter's operation, defined as "$V_{INVERTER\ INPUT}$", and *after* the inverter's operation, defined as $V_{OUT}$.

The graphs show the transient instantaneous value (red), obtained by sweeping the input voltage, $V_{IN}$, concurrently with the resistance data being fed through, while the dotted lines (yellow) show the average $V_{OUT}$ value, calculated by fixing the p-bit's $V_{IN}$ value and allowing the mimic MTJ to fluctuate over time, and averaging the output voltage over the entirety of the values produced from the resistance data available (over the 16,000 data points).

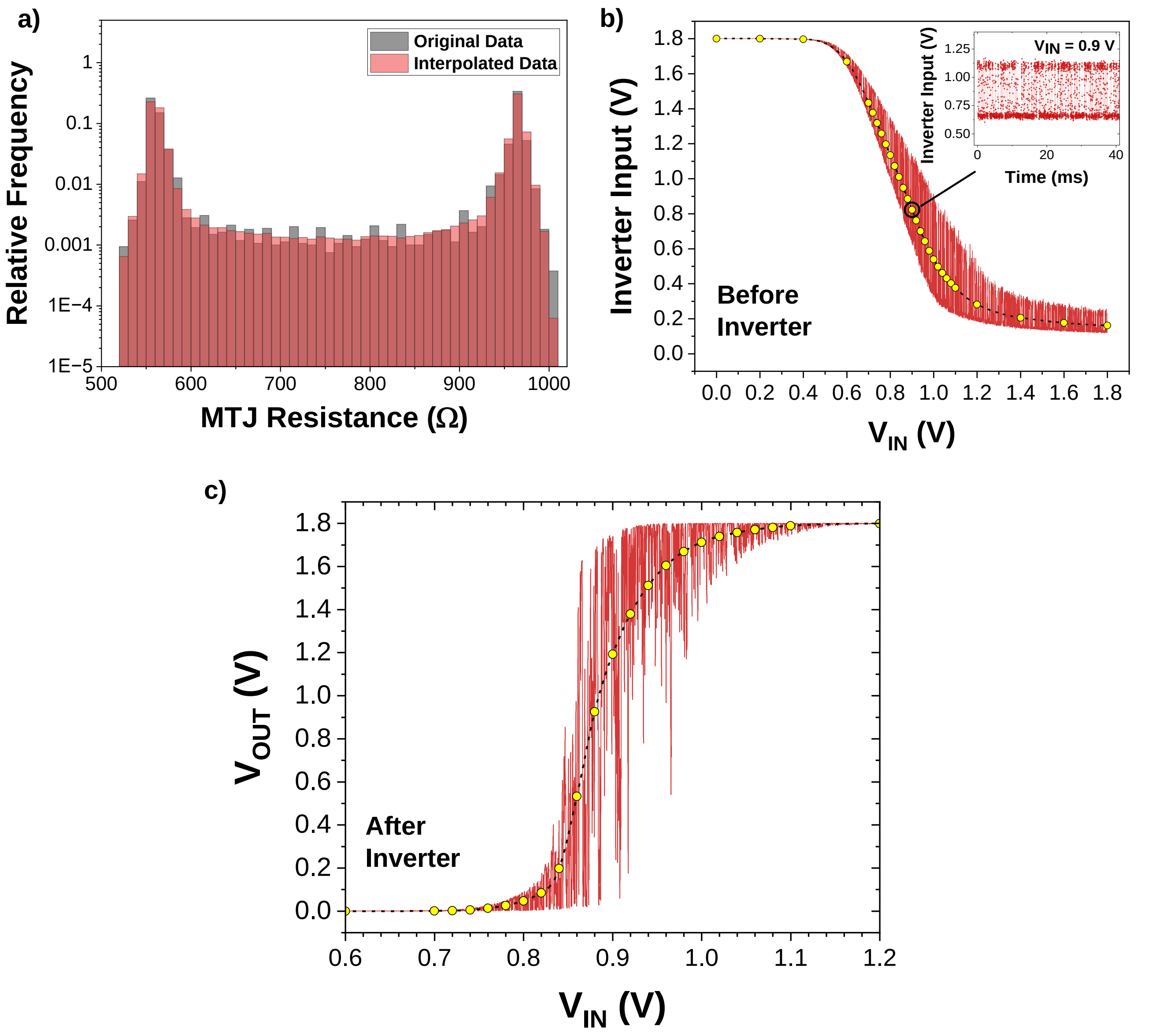

**Supplementary Figure 1: Simulation Details.** a) Plot showing the difference in the MTJ distribution when fed using the correct timestep ("Original Data"), and when using a too small timestep ("Interpolated Data"). b) The output sigmoid of the stochastic p-bit core, $V_{INVERTER\ INPUT}$, as a function of the input voltage, $V_{IN}$. The red line shows the transient instantaneous value of the voltage, while the dotted line shows the time-averaged value. c) A plot of the full p-bit's output, after the inverter's thresholding and amplification. The value of $V_{OUT}$ is pinned to 0V for $V_{IN}$ below 0.6V, and to 1.8V for $V_{IN}$ above 1.2V, and so is not shown here.

## Supplementary Information 2 – Measurement details & further characterization of Stochastic MTJs & NIST Randomness Tests

All electrical measurements are conducted in a Lakeshore probe station equipped with an in-plane magnetic field. The characterization of the transistors, and the experimental demonstration of the core of the stochastic p-bit (shown in Section 3 of the main text), are achieved using DC measurement techniques with an Agilent HP 4155B Parameter Analyzer.

The measurements of the stochastic MTJs presented in this work are performed using AC lock-in techniques, with a SR830 Lock-in Amplifier (LIA) and a Keithley 2400, in a four-point configuration to directly measure the properties of the MTJ pillar.

Supplementary Figure 2(a) shows the time-series resistance fluctuations for the MTJ device used in the 1T-1MTJ p-bit core as a function of increasing magnetic field. At an external field of -18mT, the device “spends” most of its time in the high-resistance AP-state while at -12mT, the device is predominantly in the low-resistance P-state. The dwell-times in the AP- and P-state are plotted as a function of external magnetic field, shown in Supplementary Figure 2(b), and used to identify the field at which the MTJ spends an equal amount of time in the P- and AP-state (-16mT), defined as the 50-50 point. The slight difference in tunability of the P- and AP-dwell-times by the external magnetic field, attributed to an experimental artifact of this measurement, does not appear in similar measurements of other stochastic devices, and is not analyzed further.

Aside from the tunability, a second requirement of the p-bit core for which the MTJ is used is that the stochasticity must be random. To quantify the quality of randomness, the output of these fluctuating MTJs (at the 50-50 point) are binarized (Supplementary Figure 2(c)) and put through the NIST suite of randomness tests[1,2].

Supplementary Figure 2(d) shows the results of these tests, where a p-value greater than 0.01 denotes a success, for the binary string attained when the output is sampled approximately every dwell time (~0.6s). The results show that when biased at this 50-50 point, the MTJ is indeed a source of true randomness and is suitable as the stochastic source in the p-bit core.

This is a key demonstration that highlights how low-barrier nanomagnets, coupled with MTJs for read-out, can form low-cost and compact random number generators that fluctuate naturally from ambient thermal energy, providing a platform that is especially suitable for p-bit applications.

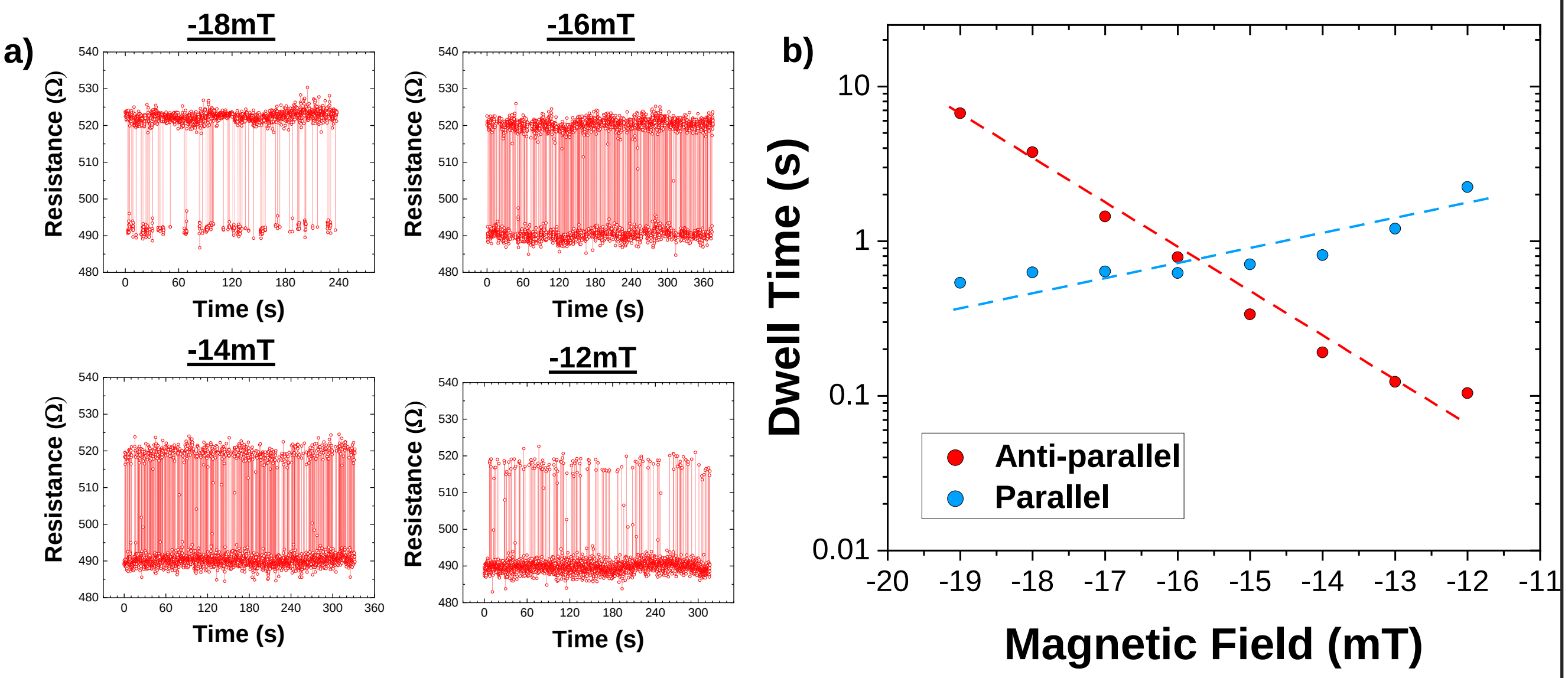

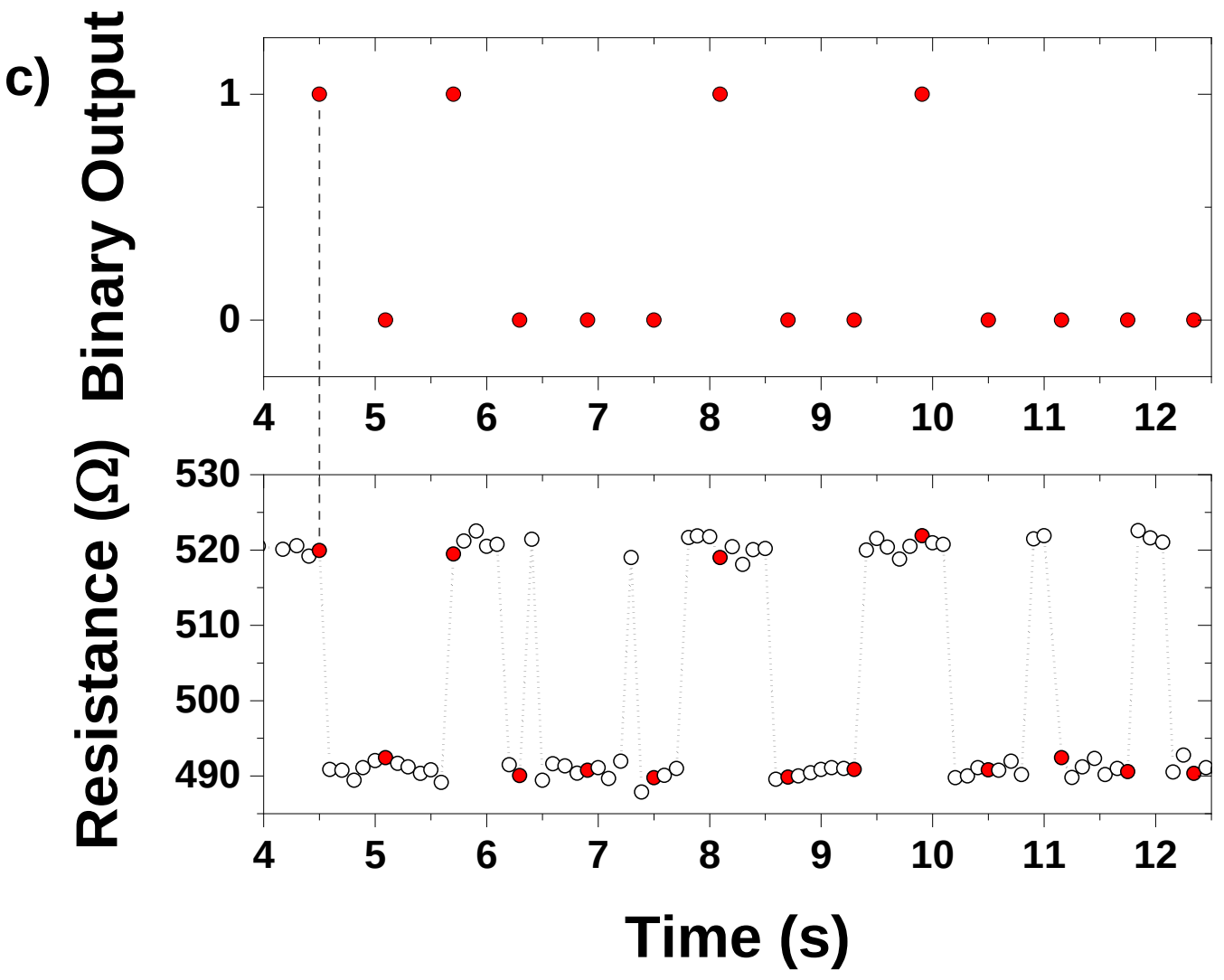

d)

| Test | Passed if p-value > 0.01 |
|---|---|
| Frequency | 0.020 ✓ |
| Block Frequency | 0.066 ✓ |
| Runs | 0.013 ✓ |
| Longest Runs | 0.021 ✓ |
| Discrete Fourier Transform | 0.303 ✓ |
| Non-overlapping Template | 0.112 ✓ |
| Serial Test | 0.683, 0.249 ✓ |
| Approximate Entropy | 1.0 ✓ |
| Cumulative Sums | 0.039 ✓ |

**Supplementary Figure 2: Electrical characterization of stochastic MTJs.** a) Time series data of stochastic resistance fluctuations at varying external in-plane field to show tunable stochasticity. b) AP- and P- dwell times plotted as a function of magnetic field to find the 50-50 point of this MTJ. c) Process showing the binarization of the resistance fluctuations to generate the bitstream for the NIST Randomness test. d) Results from the NIST randomness tests, showing the output of the MTJ is indeed truly random.

## Supplementary Information 3 – Spin Transfer Torque & Current Density Experiments

This section presents measurements on how large drive voltages may impact p-bit performance, and further explores the impact of large currents, through the resulting Spin Transfer Torque (STT) effect, on the in-plane MTJs used in the experimental demonstration of the stochastic p-bit core.

To this end, the resistance fluctuations of the MTJ used in the p-bit core demonstration are measured in a two-point configuration, while under an external field of -16mT (to bias the MTJ at the 50-50 point). This two-point configuration includes the additional resistance of the current path from the Ta channel between the contact pads and the MTJ pillar, and better mimics the actual resistance of the MTJ branch of the p-bit core.

Supplementary Figure 3(a) displays the measured values of the fluctuation's high-resistance AP-state and low-resistance P-state as a function of applied DC voltage, and illustrates that as increasingly large biases are applied across the MTJ, the base resistance degrades. This is likely because of degradation of the thin MgO layer as the stress current is increased for increasing bias voltages[3]. Reverting to a lower voltage once the MgO layer is damaged also does not fully recover performance, but results in a smaller base resistance. This is a key reason for why a $V_D$ of 200mV was chosen for the experimental demonstration in Section 3: the low RA-product of the thin MgO layer ($\sim 7\ \Omega\mu\mathrm{m}^2$) results in large current densities (>$10^6$ A/cm$^2$ for voltages above 200mV) that pose an increased risk of barrier breakdown during extended operation.

Indeed, for interconnected p-bits, parity is desired between the $V_D$ and $V_{IN}$ range such that $V_{OUT}$ can be used to drive the next p-bit in the network. For this to be achieved, and to ensure current densities are not too large to pose a risk of barrier breakdown, it is found that MTJ area-scaling is insufficient to reach the larger resistances required if the MgO thickness is too thin and the RA-product of the resulting MTJ is too small. Increasing the base resistance of the MTJ to the $10^4\ \Omega$ range will be required for future generations of unstable MTJs to achieve the operation of p-bits at the voltage range desired.

Aside from the impact on the device resistance, the impact of large current densities through STT on the MTJ's resistance fluctuations are also considered. Supplementary Figure 3(b) shows the calculated current density as well as the probability of the high-resistance AP-state for each increasing DC voltage. This quantity is calculated by binarizing the output of the resistance fluctuations, with '0' representing the low-resistance P-state and '1' representing the high-resistance AP-state, and averaging over the binary bitstream collected. Supplementary Figure 3(c) shows the time series data of the resistance fluctuations, as well as the binary output used for this operation, at applied voltages of 200mV, 400 mV and 1V.

It should be noted that the relatively slow dwell time ($\sim 0.7$s) of the device and the small sample time of measurement yields a probability-curve in Supplementary Figure 3(b) that deviates from the expected smooth sigmoid when studying the impact of STT on MTJ resistance fluctuations. Regardless, one can see that it requires a bias of 400 mV before a significant pinning effect on the fluctuations of this device is observed. This corresponds to a current density of $\sim 2 \times 10^6$ A/cm$^2$ and is consistent with the values found by others when working with the impact of STT-currents on in-plane MTJs of similar resistance[4]. This result also highlights that the 200mV chosen for the $V_D$ in the experimental demonstration in Section 3 yields a current density that *does not* cause pinning issues and does not unduly influence the MTJ's random fluctuations, preserving its integrity as a true random number generator.

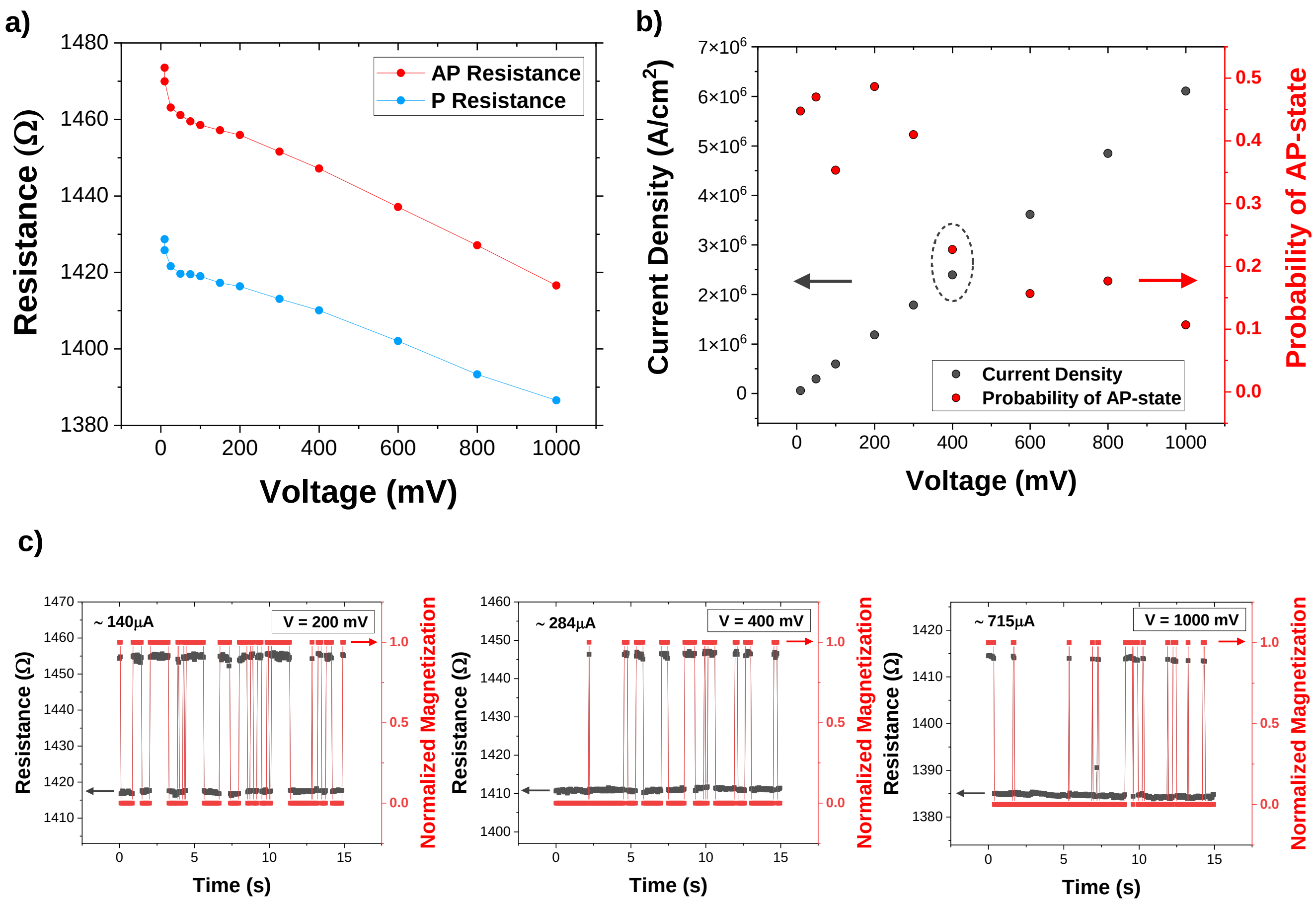

**Supplementary Figure 3: Current Density & Spin Transfer Torque (STT) experiments.** a) Plot showing the AP- and P-resistance levels as a function of applied DC voltage across the MTJ. b) The associated current density (left axis) and the calculated probability of the AP-state (right axis) for the resistance fluctuations plotted as a function of the applied DC bias across the MTJ. c) A series of graphs that shows the time series resistance fluctuations at the 50-50 point for increasing DC bias to show the impact of STT on the MTJ's fluctuations. Note that it is only at 400mV that the pinning effect to the P-state starts to be observed.

## Supplementary Information 4 – Dual-gated FETs & Threshold Voltage Control

To form networks of correlated p-bits, the output of one p-bit should be in the same voltage range as the input of the next p-bit. This means that after the inverter's operation, it is necessary for the output to be able to achieve near 0 to $V_D$ fluctuations at a specific input voltage range, and for the output sigmoid to be centered around $V_D/2$ at the input.

The 1T-1MTJ design for the p-bit core demonstrated in Section 3 utilized a back-gated $MoS_2$ transistor that, although displaying a narrow distribution of threshold voltage, $V_{TH}$, when pristine, was seen to suffer degradation through an aging effect that saw a leftward shift in the resulting $V_{TH}$ in the final integrated demonstration.

To address this, a double-gate design has been explored to offer additional control over threshold voltage and to allow a rightward shift of the threshold voltage, which controls the centroid of the sigmoidal output

The fabrication process involves first creating the back-gated $MoS_2$ devices, as described in the main body of the text, after which an Al seeding layer (~1nm) is grown and an $AlO_x$ layer is formed. Atomic Layer Deposition (ALD) of the top $HfO_2$ dielectric layer (11nm) is then performed at 90˚C after which the top-gate Au contacts are deposited. Supplementary Figure 4(a) shows a cross-section schematic of this double-gated $MoS_2$ FET. The resulting FETs are interconnected to resistance-matched MTJs across two probe-stations (Supplementary Figure 4(b)), to form an example of an off-chip p-bit core.

The transfer characteristics of a double-gated FET, shown in Supplementary Figure 4(c), demonstrate that with an increasingly negative top-gate voltage, the Fermi level is shifted towards the valence band which induces the rightward shift in threshold voltage, towards more positive back-gate voltages.

As stochastic fluctuations were not necessary for this demonstration of a double-gated p-bit core, a stable in-plane MTJ of circular design, 200nm in diameter, was chosen. The resistance vs magnetic field behavior for this device is shown in Supplementary Figure 4(d). The lack of shape anisotropy means that the magnetic moments are not defined to be colinear with the applied external field axis, and the stray field coupling of the fixed layer with the free layer ensures that at zero field, the moments are in the antiparallel high-resistance state. As the magnetic field is increased in either direction, the moments of both are brought gradually into alignment with the externally applied field, yielding a low-resistance parallel-state at high fields.

Supplementary Figure 4(e) shows the output for this double-gated p-bit core, measured at the drain of the FET, $V_{INVERTER\ INPUT}$, at different set top-gate voltages while the back-gate voltage ($V_{IN}$) is swept from -1.7V to +1.5V. The $V_D$ is limited to 400mV to minimize the risk of oxide breakdown in the MTJ. Overall, it is seen that with application of increasingly negative voltages to the top-gate, the threshold voltage, and the resulting p-bit core output sigmoid, can be shifted to the right. Indeed, for $V_{TG}$ = -2.5V, the output sigmoid is well-centered around 0V and may be pinned with the noise margin at the output-high and output-low values.

This demonstrates that a dual-gated FET design is a viable pathway to achieving threshold voltage control, resulting in a well-centered sigmoidal output that is suitable for the next-step of inverter amplification, and further offering an additional mode of control for in-situ adjustments of devices in correlated p-bit networks.

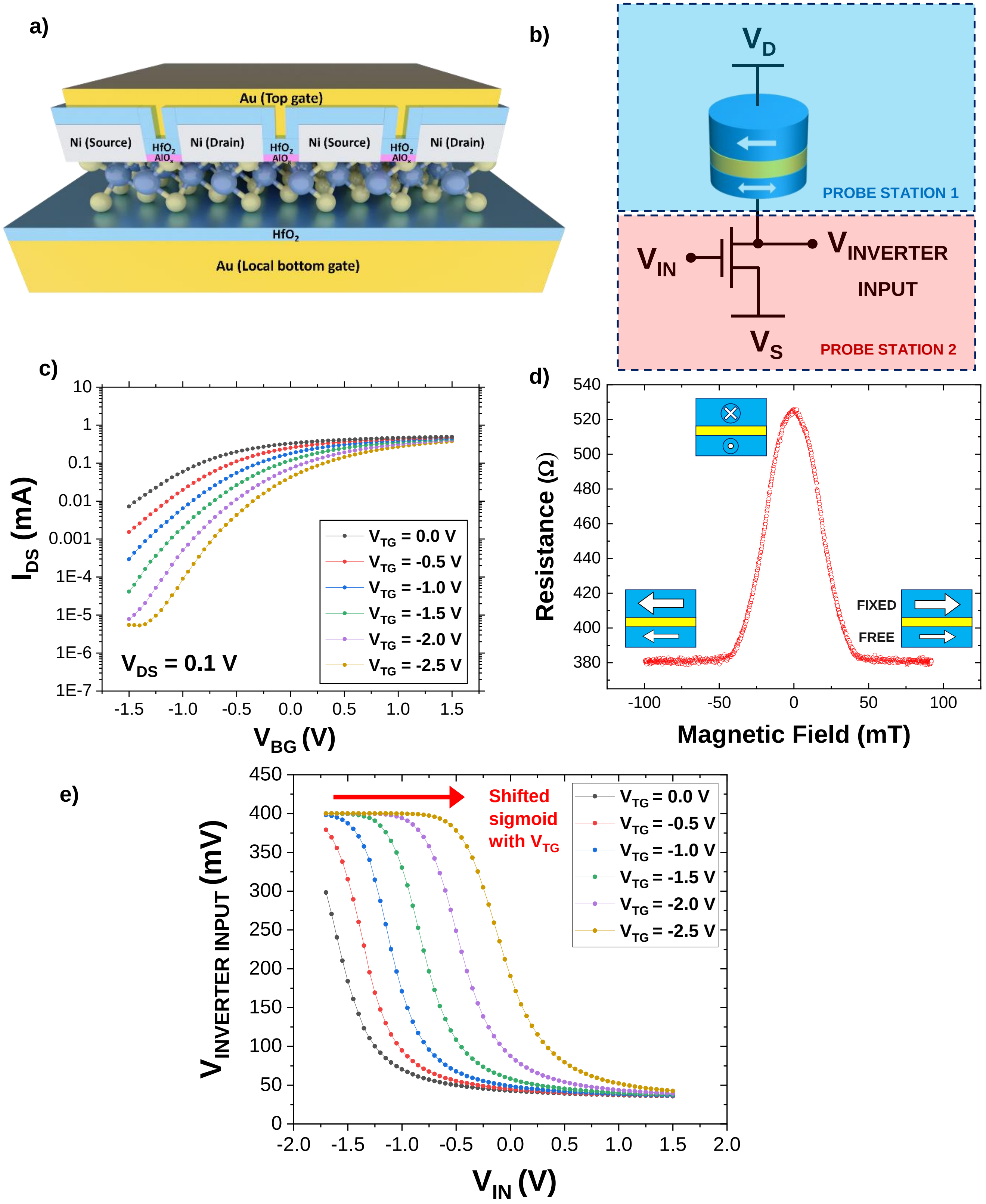

**Supplementary Figure 4: Control of threshold voltage and sigmoid-center using a double-gated FET.** a) Cross-section of the double-gated $MoS_2$ FET structure. b) Schematic of the integrated p-bit core, showing the split of devices across two probe-stations. c) FET transfer characteristics where a rightward shift in $V_{TH}$ at increasingly negative top-gate voltages is observed. d) Four-point measurement of the MTJ pillar showing the major loop of the stable in-plane MTJ device used for this demonstration. e) Sigmoidal output for this integrated double-gated p-bit core, demonstrating that the sigmoid may be shifted with application of a top-gate voltage.

## Supplementary Information 5 – Further demonstration of 1T-1MTJ p-bit core using well-resistance matched MTJs & FETs

This section covers examples of better resistance-matched MTJ & FETs and aims to further highlight how aspects of each can influence the sigmoidal output of this p-bit core structure. The p-bit core devices presented here are examples of a well-resistance matched device and a device that utilizes a fast MTJ; both devices are integrated across two probe stations in an off-chip structure, as shown in Supplementary Figure 5(a).

Supplementary Figure 5(b) shows the FET transfer characteristics, and Supplementary Figure 5(c) and 5(d) show the MTJ characteristics, for the slow but well-resistance matched p-bit core while Supplementary Figure 6(a) and Supplementary Figure 6(b) and 6(c) show the FET, and MTJ, characteristics for the second p-bit core, respectively. The minor loops in Supplementary Figure 5(c) (and Supplementary Figure 6(b)) are measured using four-point lock-in techniques across just the MTJ pillar, while the time series resistance fluctuation measurements in Supplementary Figure 5(d) (and Supplementary Figure 6(c)) are two-point measurements, measured at the MTJ's 50-50 point, that include the additional Ta channel resistance.

First, it is noted that both transistors have a $V_{TH}$ close to zero, without top-gate control, and this results in well-centered sigmoidal outputs.

Secondly, consider the resistance-matching between the MTJ and FET for the two p-bit core devices. Supplementary Figure 5(e) shows the output of the first, well-resistance matched, device where in the off-state the FET is many orders of magnitude more resistive than the MTJ, allowing for the output to be easily pinned to $V_D$ at low $V_{IN}$. In the on-state, the FET resistance reaches the $10^2$ Ω range with the high current-drive provided by the interdigitated finger design, while the MTJ has a resistance in the $10^4$ Ω range. The resulting voltage division means the output may be pinned effectively to 0 at large $V_{IN}$.

In contrast, Supplementary Figure 6(d) shows the output of a p-bit core where the resistance matching is not as ideal. In the off-state, the MTJ resistance is much less than the FET resistance, allowing the output to be pinned to $V_D$ at low $V_{IN}$. However, the lower base resistance of the MTJ (in the $10^3$ Ω range) means in the on-state, where the FET resistance is in the $10^2$ Ω range, gives only a 10:1 resistance ratio that means that for large $V_{IN}$, the output does not approach 0 but rather $V_D/11 \approx 27$mV. This is still acceptable with an inverter that has a suitable noise margin but is an example of a less-ideally resistance-matched p-bit core.

Another difference between the two outputs is in the speed of fluctuation of the MTJ used in the corresponding p-bit cores. The non-ideal case here are the fluctuations from the slow MTJ, shown in Supplementary Figure 5(d), where the bimodal resistance distribution manifests as a very bimodal distribution in the $V_{INVERTER\ INPUT}$, shown in the inset of Supplementary Figure 5(e). As described in Section 4 of the main text, this can cause problematic plateaus in the overall p-bit's output.

In contrast, the faster MTJ of Supplementary Figure 6(c) has a much more continuous distribution, where the resistance values in-between the bounding low- and high-resistance states of the fluctuations are sampled more evenly, and the $V_{INVERTER\ INPUT}$ values (shown in inset of Supplementary Figure 6(d)) reflect this. This is crucial for attaining a p-bit output sigmoid that is devoid of the non-ideal plateaus described previously.

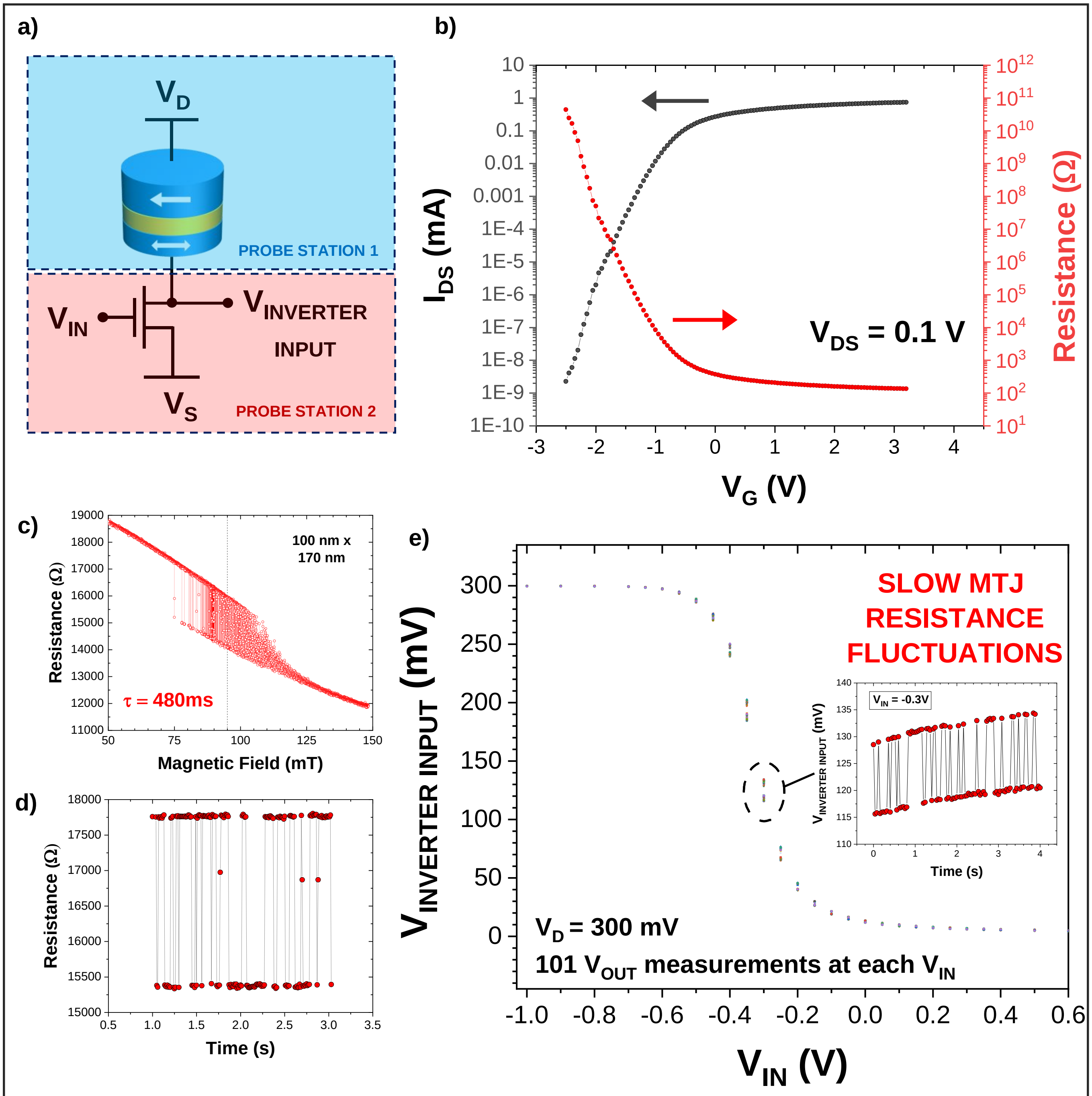

**Supplementary Figure 5: Measurements of a well-resistance matched p-bit core.** a) Schematic of the p-bit core, showing the split of devices across the two probe-stations. b) FET transfer characteristics, with the drain current (left-axis) and the calculated FET resistance (right-axis) plotted as a function of gate voltage, $V_G$, at a drain voltage of 0.1V. c) Four-point measurement of the MTJ pillar showing the minor loop and the 50-50 point. d) Two-point time-series measurements of the MTJ to demonstrate the bimodal resistance fluctuations when biased at the 50-50 point. e) Output for the p-bit core made with the bimodal MTJ, showing good pinning at $V_D$ and 0 for low and high $V_{IN}$, respectively, but a bimodal voltage distribution at the $V_{INVERTER\ INPUT}$.

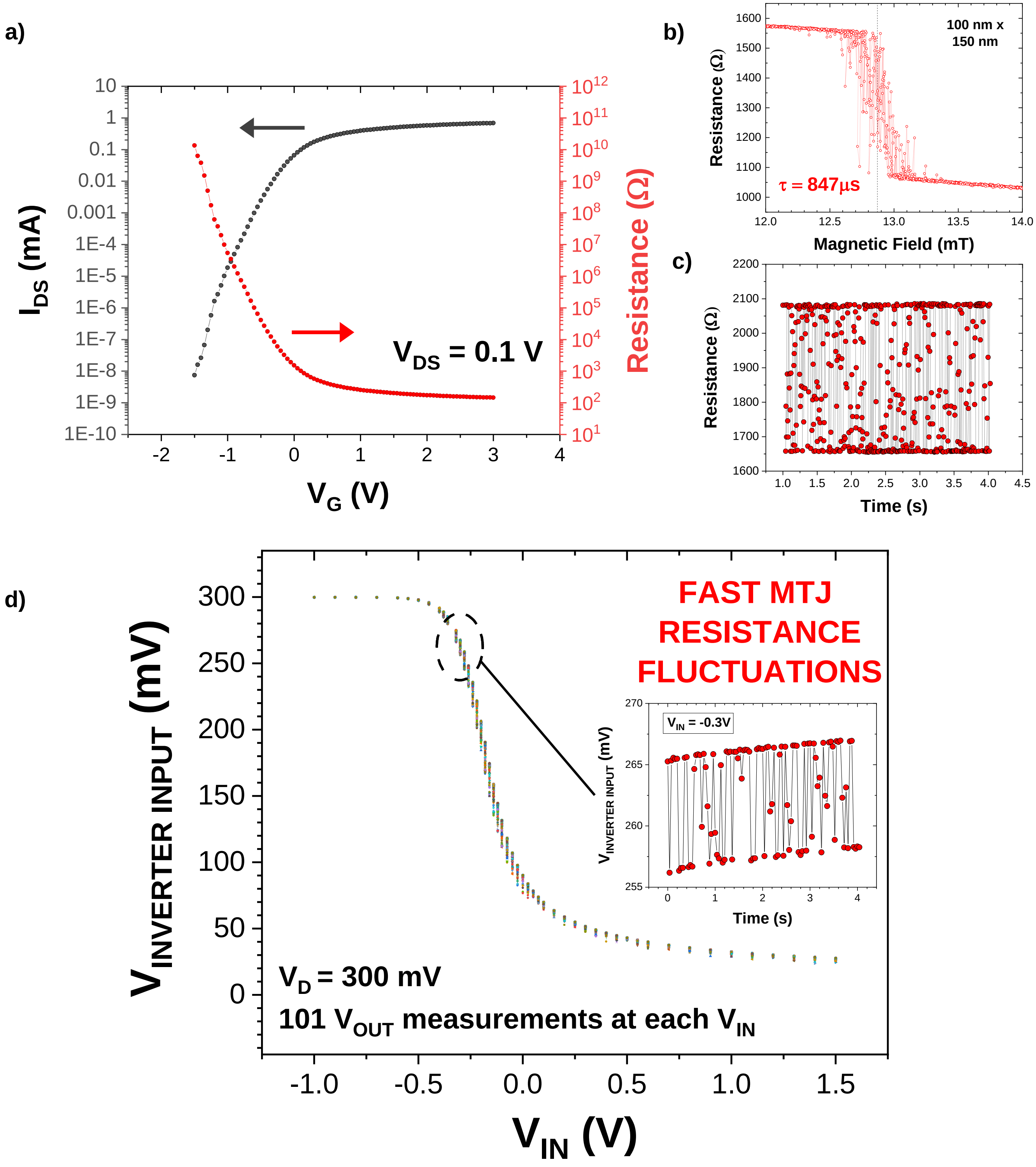

**Supplementary Figure 6: Measurements of a fast MTJ p-bit core.** a) FET transfer characteristics, showing the drain current (left-axis) and the calculated FET resistance (right-axis) at a drain voltage of 0.1V. b) Four-point measurement of the MTJ pillar displaying the minor loop and highlighting the 50-50 point. c) Two-point time-series measurements of the MTJ to demonstrate the continuous resistance fluctuations when biased at the 50-50 point. d) Output for the p-bit core made with the continuous MTJ, showing good pinning at $V_D$ for low $V_{IN}$ but the inability to pin to 0 at high $V_{IN}$. However, the continuous resistance distribution of the MTJ manifests as a continuous voltage distribution at the $V_{INVERTER\ INPUT}$.

## Supplementary Information 6 – TMR Scaling for simulation input

To understand how the Tunnel Magnetoresistance (TMR) coefficient of a MTJ affects the output of a p-bit, three different resistance ranges are chosen to yield TMR coefficients of ~15%, ~80% and ~300%. The input data is taken from time series measurements of a real stochastic MTJ device, fluctuating at its 50-50 point, and exhibiting a dwell time of 847us. The raw input is then normalized between the range of [0,100] and scaled to the three TMR ranges described previously. The resulting time-series outputs are shown in Supplementary Figure 7(a), where each output consists of 16,000 data points.

To verify that this scaling is valid, statistics are presented for 23 actual stochastic devices to map any possible correlations that could impact the validity of the TMR scaling approach employed in this work, and to see if the magnitude of the resistance fluctuations influences the dwell time.

Supplementary Figure 7(b) shows the dwell time plotted against the resistance range over which fluctuations are observed, $R_{AP} – R_P$, while Supplementary Figure 7(c) shows the dwell time plotted against the TMR of the stochastic fluctuations. As the plots show, there is no correlation between the dwell time and either of these quantities and so the TMR scaling can be performed without changing other key properties of the stochastic MTJ.

The resulting TMR distributions are shown in Supplementary Figure 7(d). It should be noted that the scaling is done to ensure the midpoint for each device is the same, i.e. 750Ω, so that the same transistor and inverter may be used in all three cases without concerns regarding resistance matching, and to preserve the overall distribution of states.

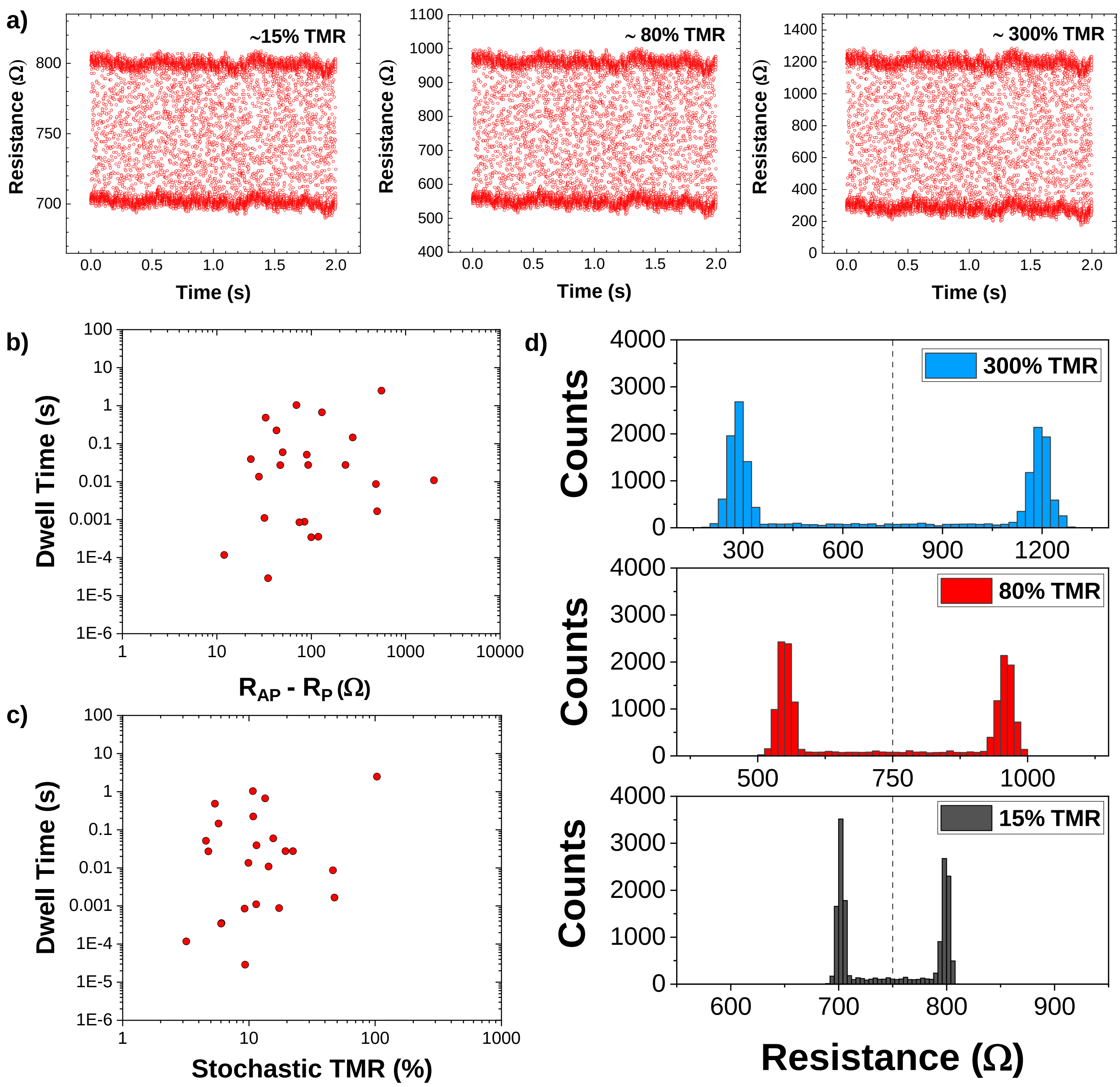

**Supplementary Figure 7: Details on TMR scaling for simulation input.** a) Time series data for the ~15% TMR, ~80% TMR and ~300% TMR device constructed from real experimental data of a stochastic MTJ. b) Plot displaying statistics for 23 stochastic MTJ devices, showing how the dwell time does not correlate with the size of the stochastic resistance fluctuation ($R_{AP} – R_P$). c) Plot demonstrating how the measured dwell time has no correlation with the MTJ's TMR. This suggests the scaling is a valid approach without introducing other variables that might impact the distribution/speed of fluctuation. d) Histograms for the three devices of increasing TMR used in the simulation, highlighting how the midpoint is kept at the same value so the same resistance-matched transistor and inverter may be used across all three p-bits.

## Supplementary Information 7 – Explanation of voltage distribution for impact of TMR on p-bit's output

This section aims to further clarify the nature of the histograms showing the distribution of voltages at the inverter's input for the p-bits made with MTJs of different TMRs (Figure 4(c) of the main text). Three cases were considered in particular: for $V_{IN}$ = 0.8V, $V_{IN}$ = 0.9V and $V_{IN}$ = 0.98V, with a focus here on explaining the broadening of the distributions, as well as the change in shape and symmetry of the distributions, as $V_{IN}$ increases between these three cases.

A key point to note is that the MTJs in this p-bit design are primarily "noisy resistors", and only provide the stochastic behavior. As a result, their resistance values are confined to fluctuate between the $R_{AP}$ and $R_P$ values, and do not change as a function of $V_{IN}$ (Supplementary Figure 8(a)). The tunability is obtained by modulating the gate voltage of the transistor that is in series with the MTJ, such that the voltage at the inverter's input may be tuned according to the relative resistance between, and voltage drop across, the MTJ and FET.

For a fixed input voltage ($V_{IN}$), the stochastic resistance fluctuations of the MTJ lead to a fluctuating drain voltage for the FET, resulting in a fluctuating current through the MTJ + FET (as shown in Supplementary Figure 8(b)).

For small $V_{IN}$, where the transistor is in the off-state and the resistance is much larger than the MTJ resistance, this effect is minimal, and the fluctuations are suppressed. However, as the MTJ and FET resistance are brought into similar ranges with increasing $V_{IN}$, the impact of the MTJ's fluctuations, and the relative change in the drain voltage of the FET, are much larger. This occurs around $V_{IN} \approx 0.8$V, as seen in Supplementary Figure 8(c) and 8(d), and helps explain the widening of the distributions as $V_{IN}$ is further increased.

This broadening of the $V_{INVERTER\ INPUT}$ values is also seen in Supplementary Figures 8(e) and 8(f), where bigger swings in voltage drop are first observed at $V_{IN}$ where the MTJ and FET resistances are of a similar magnitude.

These graphs, which show the output of the stochastic p-bit core before inverter amplification, also explain the apparent asymmetry in the 300%-TMR distributions at $V_{IN}$ = 0.98V. Note that the lower arm of the distribution corresponds to the high-resistance state of the 300%-TMR MTJ. At $V_{IN}$ = 0.98V, the ratio of MTJ resistance to FET resistance (in the low $10^2\ \Omega$ range) is such that the 300%-TMR MTJ is nearly always at a much larger resistance than the FET, resulting in more $V_{INVERTER\ INPUT}$ values that are small and an asymmetrical distribution.

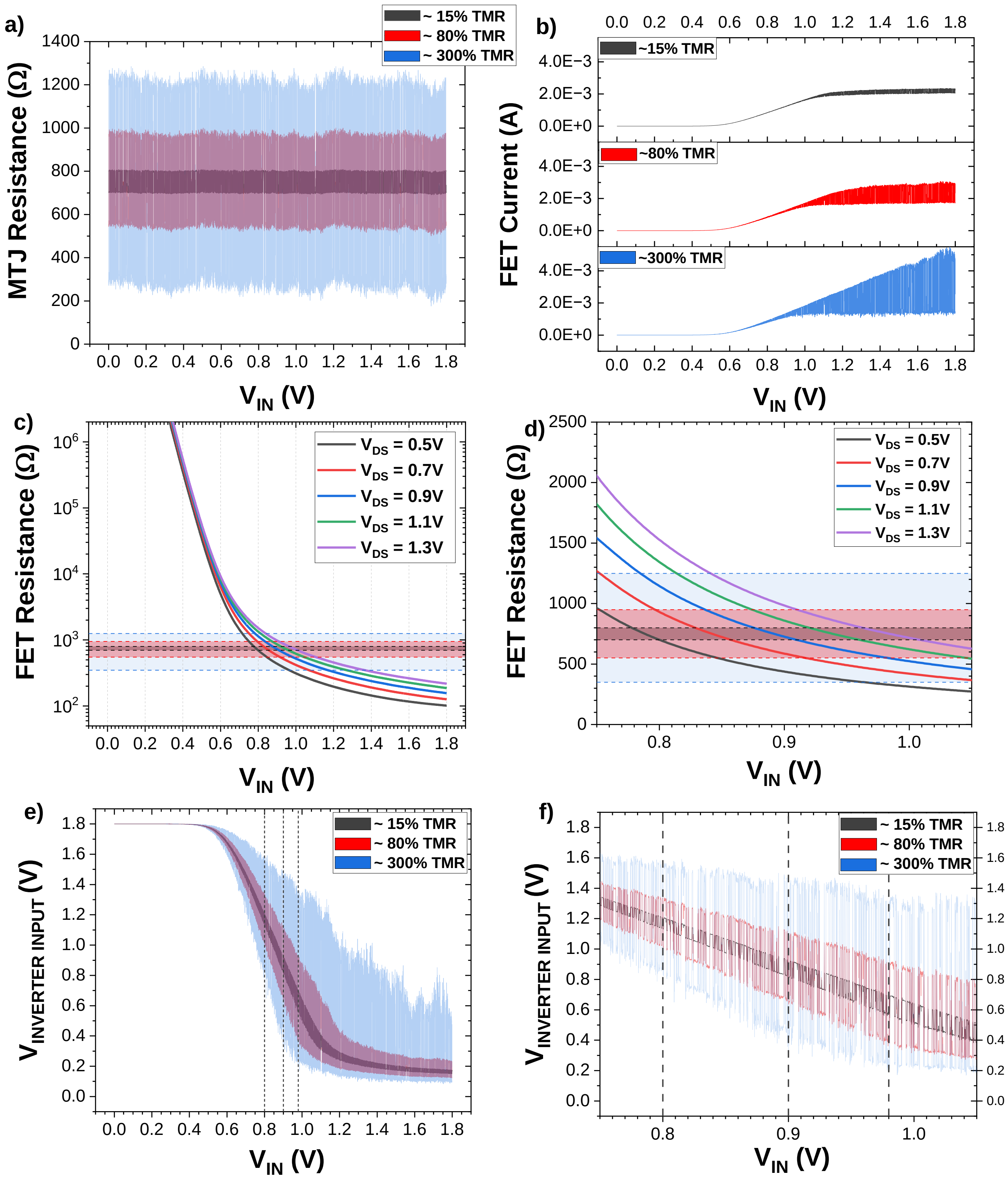

**Supplementary Figure 8: Further explanation of the voltage distribution histograms for the impact of TMR on the p-bit's output.** a) Plot showing how the resistance fluctuations of the three TMR-devices are invariant with $V_{IN}$. b) Graph plotting the NMOS FET current against the input voltage, demonstrating how the resistance fluctuations in the MTJ manifest as fluctuations in the current through the FET. c) Plot of the FET resistance, calculated at different $V_{DS}$, along with the MTJ resistance (colored bands) for the full $V_{IN}$ range and d) for the narrow range shown in the histograms in Section 5 of the main text. e) Plot of the voltage at the $V_{INVERTER\ INPUT}$ point for the three TMR devices as a function of the full $V_{IN}$ range, and f) for the narrow range of $V_{IN}$ from which the histograms are taken.

## Supplementary Information 8 – The Distribution Factor

The 'distribution factor' is a quantity that is introduced as a measure of how bimodal, or continuous, an MTJ's resistance fluctuations are. It is calculated by taking the resistance fluctuations of an MTJ, normalizing the output to the range [0,100], and creating a histogram distribution of the resulting normalized output. Care is taken to ensure the same size of data set is used for each stochastic device (16,000 resistance data points for each distribution), and that equal bin sizes are used (to make the comparison of counts statistically useful).

The counts in the 8 'edge'-state bins (which represent the counts in the P- and AP-resistance states) are divided by the counts in the 8 'middle'-state bins (which represent the counts in-between the bounding P- and AP-state) to give a number defined as the 'distribution factor'. A value that is very large indicates a very bimodal distribution, whereby the number of edge states are far more numerous than the number of middle states, whereas a small number indicates a more continuous distribution where the number of edge and middle states are similar. An ideal 'uniform' distribution, where the number of edge and middle states is identical, would be represented by a distribution factor of 1.

A key finding of this work is that there appears to be a correlation between this distribution factor, or the distribution of resistance states, and the dwell time of the stochastic device. A faster device is observed to have a more continuous distribution, with more middle states, and a correspondingly smaller distribution factor. This is important as a more continuous distribution is shown to be desirable, with the resulting p-bit output having fewer non-ideal plateaus the more continuous the distribution is.

One potential worry is that this quantity is a function of the sampling time of measurement. Faster devices require faster sampling times (shorter $\Delta\tau$ between measurement points) to accurately capture their fluctuations, and this could account for the correspondingly smaller distribution factor as the number of middle states would be inflated.

To account for this, the same stochastic device is measured with three different sampling times (shown in Supplementary Figure 9(a)): with sampling times of $\Delta\tau = 3.1\mu s$, $\Delta\tau = 15.5\mu s$, and $\Delta\tau = 77.5\mu s$. Supplementary Figure 9(b) shows the number of counts of middle states (left axis), and the calculated distribution factor (right axis) for each of these sampling times. As expected, the shortest sampling time gives a much larger count of the middle states compared to the slowest sampling time, but the distribution factor is observed to be rather invariant as both the edge and middle states are similarly well-sampled. This proves that it is valid to compare the distribution factor measurements for devices across different dwell time, and sampling time, regimes.

To further validate this quantity, and to see if other correlations might arise, the distribution factor is plotted against other properties of a stochastic MTJ. This includes plotting the distribution factor against the average resistance of the stochastic fluctuation (Supplementary Figure 9(c)), against the magnitude of the stochastic fluctuations (Supplementary Figure 9(d)), and finally, against the bias field where the stochastic fluctuations occur (Supplementary Figure 9(e)). As can be seen in each of these cases, there is no clear correlation between the distribution factor and any other property of the stochastic MTJ other than its unique dependence on the dwell time of the device.

One possible explanation for this is that this distribution factor, which compares the number of counts of the edge states to middle states in the resistance distribution of an MTJ's stochastic fluctuations, is representative of the amount of time the MTJ's free layer spends in the P- or AP-state (the edge-state counts) compared to the amount of time the free layer spends in transitioning between them (the middle-state counts). This transition time is dependent on material properties of the MTJ stack layers, such as the perpendicular effective anisotropy field and the damping factor, and for in-plane MTJs is theorized to be in the range of approximately 1-10ns[5]. Therefore, the smaller the energy barrier, the smaller the dwell time in the P- or AP-state, while the transition time is relatively unaffected (with change in the energy barrier size). Thus, the smaller the dwell time, the fewer the edge states relative to middle states, which correlates to a smaller distribution factor.

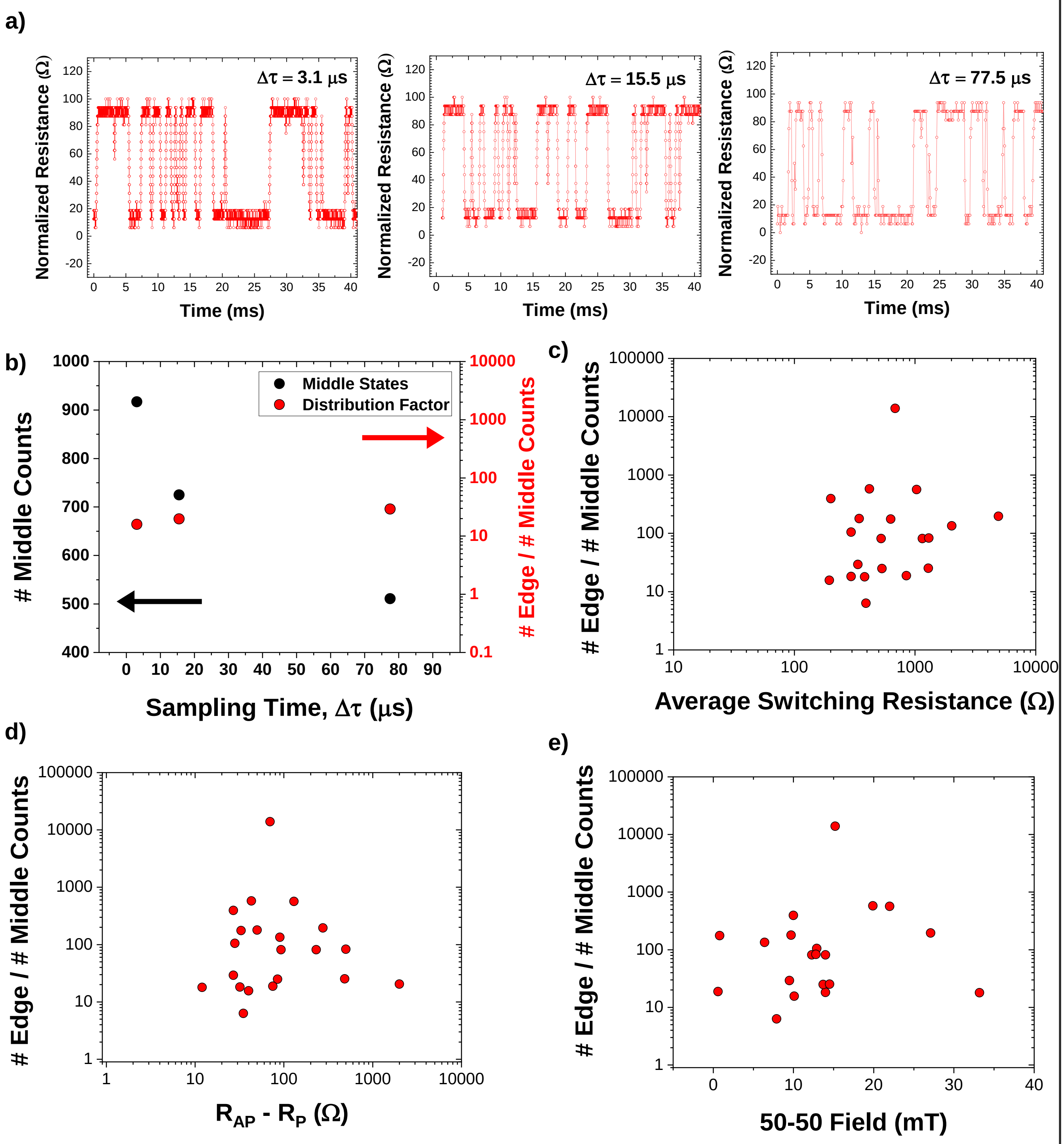

**Supplementary Figure 9: Supporting information for the 'Distribution Factor'.** a) A series of plots displaying the time-series data taken at the 50-50 point for the same MTJ, but with an increasing sampling time, $\Delta\tau$, between data points. B) Plot showing how the increasing sampling time does lead to a decreasing number of 'middle count' states (left axis), but the 'Distribution Factor' is relatively unaffected (right axis). Graphs plotting statistics from 23 stochastic devices demonstrating how the 'Distribution Factor' of these devices show no correlation with the c) average resistance, d) the magnitude of resistance fluctuations or e) where the stochastic switching is observed to occur (in terms of the externally applied magnetic field). The only correlation observed for this quantity is with the device's dwell time, as presented in the main text.

## Supplementary Information References